\documentclass[twocolumn]{autart}
\usepackage{amssymb}
\usepackage{amsmath}
\usepackage[round]{natbib} 
\usepackage{color}
\usepackage{graphics}
\usepackage{graphicx}
\usepackage[tight,footnotesize]{subfigure}
\usepackage{multirow}
\usepackage{algorithm,algorithmic}
\usepackage{savesym}
\usepackage[justification=centering]{caption}
\savesymbol{AND}

\usepackage{tikz}
\usetikzlibrary{fit,shapes.misc}

\def \kkk{\color{black}}

\def \R{\mathbb{R}}

\def \E{\mathcal{E}}
\def \V{\mathcal{V}}

\def \N{\mathcal{N}}

\def \B{\mathcal{B}}

\def \O{\mathcal{O}}
\def \P{\mathcal{P}}

\def \U{\mathcal{U}}
\def \X{\mathcal{X}}

\def \G{\Gamma}

\def \a{\alpha}
\def \o{\omega}
\def \e{\epsilon}

\newtheorem{theorem}{Theorem}
\newtheorem{remark}{Remark}
\newtheorem{lemma}{Lemma}

\journal{Automatica}

\begin{document}
\maketitle
\thispagestyle{empty}
\pagestyle{empty}	

\begin{frontmatter}

\title{Consensus of Multi-agent System via Constrained Invariant Set of a class of Unstable System}	


\author[AddA]{Chong Jin Ong}\ead{mpeongcj@nus.edu.sg},
\author[AddA]{Bonan Hou}

\address[AddA]{Department of Mechanical Engineering, National University of Singapore, 117576, Singapore}


\begin{abstract}
This work shows an approach to achieve output consensus among heterogeneous agents in a multi-agent environment where each agent is subject to input constraints. The communication among agents is described by a time-varying  directed/undirected graph.  The approach is based on the well-known Internal Model Principle which uses an unstable reference system. One main contribution of this work is the characterization of the maximal constraint admissible invariant set (MCAI) for the combined agent-reference system. Typically, MCAI sets do not exist for unstable system. This work shows that for an important class of agent-reference system that is unstable, MCAI exists and can be computed. This MCAI set is used in a Reference Governor approach, combined with a projected consensus algorithm, to achieve output consensus of all agents while satisfying constraints of each.  Examples are provided to illustrate the approach.
\end{abstract}

\end{frontmatter}




%






\section{Introduction}

The consensus problem of Multi-agent system (MAS) refers to achieving a consensus value of the outputs of a group of agents connected via a communication network. The study of this problem has been an active area of research in the last decade and thus, has a rich literature. See for example \citep{BOO:ME10,ART:LR14,ART:NOP10,ART:SS09,ART:R08,ART:WSA11} and references therein. Since all physical systems have constraints, achieving consensus under constraints is clearly an important consideration. However, most research effort in MAS considers agents that are constraint-free except for a notable few.  These include \cite{ART:R08} where each agent is a double-integrator system under a fixed directed communication network; \cite{ART:YMDJ14} where all agents are described by one general linear system in a fixed communication network; \cite{ART:WYG14} where each agent is either Lyapunov stable or a double integrator and all agents are homogeneous;  \cite{ART:MZL13} where agents are homogeneous general linear system under a leader-follower switching network.

This work differs from those mentioned above in that the proposed approach is for general linear system under a switching network where agents are heterogeneous subject to constraints on their inputs and following trajectories generated from a class of unstable reference models. Such trajectories are common in formation control of MAS where the positions of the agents are affine functions of time. However, the more significant difference is in the solution approach - it is based on the Maximal Constraint Admissible Invariant Set (MCAI) for the combined system of the agent and the desired reference. The use of MCAI sets for constraint satisfaction is well known  \citep{ART:GT91,ART:GO11,ART:Bla99,ART:DO12,ART:AJ18,INP:WJO19}. For example, these sets are featured extensively in Model Predictive Control \citep{ART:MRRS00,ART:T16,ART:OG06} and are an integral part of the workings of Command/Reference Governors \citep{ART:CMP04,ART:GO11,ART:GCK17,INP:KKG12,ART:KKG14}. To the best of our knowledge, all MCAI sets in the literature are obtained from systems that are stable or, at least, Lyapunov stable.
This is expected since MCAI sets do not exist for most unstable systems.
This work uses MCAI set where the reference system is a special class of unstable systems, commonly used in multi-agent applications. This MCAI set is used in the Reference Governor (RG) framework  \citep{ART:GK02,ART:GO11,ART:GCK17} to achieve pointwise-in-time constraint satisfaction for each agent while achieving output consensus among the agents. While the proposed approach is for a multi-agent system, the MCAI set can be of independent interest for a single system under the Internal Model Principle (IMP) settings, commonly known as the output regulation problem.

The study of consensus in a similar setting but without constraint has appeared \citep{ART:WSA11}. In that work, IMP conditions \citep{ART:FW76,BOO:KIF93} are shown to be both necessary and sufficient for achieving output consensus among heterogeneous agents. In the presence of constraints, the IMP conditions are only necessary.  This work shows how additional conditions can be obtained to ensure constraint satisfaction while achieving consensus. A recent work by the authors \citep{ART:ODH20} addresses the same consensus problem among heterogeneous agents with constraints but for the case where the reference system is Lyapunov stable.  As shown from section \ref{sec:Choice} onwards, the case where the reference system is unstable is considerably different from the stable case, with new formulation of the Reference Governor, characterization of the unstable MCAI sets and associated technical results and properties.

The rest of this paper is organized as follows. This section ends with a description of the notations used. Section \ref{sec:pre} describes the problem statement, reviews standard communication network and the internal model principle.
Section \ref{sec:Choice} discusses the implications of the choices of the reference system, agent dynamics and their prevalence in the multi-agent consensus settings. Section \ref{sec:main} begins with the characterization of the maximal output admissible set and its properties under the controller obtained from IMP with the unstable reference system for a single system. The overall controller with the Reference Governor is described in later part of Section \ref{sec:main} followed by related results. The multi-agent case is discussed in Section \ref{sec:multi}, followed by Section \ref{sec:global} where an approach to enlarge the domain of attraction is discussed. The performance of the approach is illustrated using an example in Section \ref{sec:num}. Conclusions are given in Section \ref{sec:con}.

The notations used in this paper are standard. Non-negative and positive integer sets are indicated by $\mathbb{Z}^+_0$ and $\mathbb{Z}^+$ respectively. Selected ranges of the integer set are $\mathbb{Z}^N=\{1,\cdots, N\}$ and $\mathbb{Z}_{\ell}^k=\{\ell, \ell+1, \cdots, k\}$ where $k > \ell$.
Similarly,  the sets of real numbers, $n$-dimensional real vectors and $n$ by $m$ real matrices are $\mathbb{R}, \mathbb{R}^n, \mathbb{R}^{n \times m}$ respectively. $I_n$ is the $n\times n$ identity matrix, $1_n$ is the $n$-column vector of all ones and $\B_n(\e):=\{x \in \mathbb{R}^n: \|x\|_2 \le \e\}$ is the $\e$-ball in $\mathbb{R}^n$; abbreviated as $I$, $1$ and $\B(\e)$ when the dimension is clear. Given sets $X,Y \subset \mathbb{R}^n$, $int(X)$ is its interior and $X \oplus Y=\{ x + y| x \in X, y \in Y\}$ is their Minkowski sum.  The transpose of matrix $M$ is $M'$. The sign of $y \in \mathbb{R}^n$ is $sgn(y)$ interpreted element wise. For a square matrix $Q$, $Q \succ (\succeq) 0$ means $Q$ is positive definite (semi-definite) and $spec(Q)$ is its set of eigenvalues. Given a set of vectors $S=\{v_i \in \mathbb{R}^n\}_{i=1}^{N}$, the span of $S$ is the set of all linear combinations of the vectors.
The $2-$ and $P$-norm of $x \in \mathbb{R}^n$ are $\|x\|_2$ and $\|x\|_P$ respectively with $\|x\|^2_P=x'Px$ for $P \succ 0$.
Given $M_i \in \mathbb{R}^{n \times m}, x_i \in \mathbb{R}^n, i \in \mathbb{Z}^N$, $M^j_i$ refers to the $j^{th}$ column of $M_i$ and $x^j_i$ is the $j^{th}$ element of $x_i$, 
and $\hat{\pmb{x}}=[ (x_1^1 x_2^1 \cdots x_N^{1}) (x_1^2 \cdots x_N^2),\cdots (x_1^n x_2^n \cdots x_N^n)]' \in \mathbb{R}^{Nn}$. Diagonal matrix is denoted as $diag\{d_1,\cdots,d_n\}$ with diagonal elements $d_i$. Additional notations are introduced when required.

\section{Preliminaries}\label{sec:pre}
This section consists of two subsections that review background materials on multi-agent systems and Internal Model Principle respectively.
\subsection{Multi-agent System}
The system considered herein is a network of $N$ discrete-time linear systems, each of which is described by
\begin{subequations} \label{eqn:system}
\begin{align}
x_i(t+1)&=A_i x_i(t) + B_i u_i(t), i \in \mathbb{Z}^N, t \in \mathbb{Z}^+_0 \label{eqn:xikplus1}\\
y_i(t)&=C_i x_i(t), i \in \mathbb{Z}^N, t \in \mathbb{Z}^+_0 \label{eqn:yi}\\
u_i(t) &\in \U_i, i \in \mathbb{Z}^N, t \in \mathbb{Z}^+_0 \label{eqn:xiui}
\end{align}
\end{subequations}
where $x_i(\cdot) \in \R^{n_i}, u_i(\cdot)\in \R^{p_i}, y_i(\cdot) \in \R^{q}$ are the state, control and output of the $i^{th}$ system, $\U_i \subset \R^{p_i}$ is the constraint set on $u$. The objective is to have $y_i, i \in \mathbb{Z}^N$ reach consensus among the $N$ system while satisfying (\ref{eqn:xiui}) at all time. 

The $N$ subsystems are connected on a time-varying directed network which is described by a weighted graph $G(t) = (\V,\E(t))$ with vertex set $\V = \{1,2,\cdots,N\}$, edge set $\E(t) \subseteq \V \times \V$ with $(i,j) \in \E(t)$ meaning $j$ being an in-neighbor of $i$ at time $t$.
Neighbors of node $i$ are $\N_i(t):=\{j\in \V:  (i,j) \in \E(t), i \neq j\}$. The communication network among the agents is described by a $N \times N$ matrix $\P(G(t))$ with elements
\begin{align} \label{eqn:Perron}
\P_{ij}(G(t))=\left\{
\begin{array}{ll}
a_{ij}(t), & \hbox{$(i,j)\in \E(t)$;} \\
1-\sum\limits_{j \in \N_i(t)} a_{ij}(t), & \hbox{$i=j$;} \\
0, & \hbox{otherwise.}
\end{array}
\right.
\end{align}
where $a_{ij}(t) \leq 1$ and, if nonzero, is greater than $\bar{a}$, for some $\bar{a}>0$.

Several assumptions are needed. These are \\
(\textbf{A1}) The graph $G(t)$ is uniformly strongly connected, or equivalently,  for every $t$, there exists a finite $T > 0$ such that the union of the graphs from $t$ to $t+T$ given by $\bar{G}(t,t+T):= G(\V, \cup_{k \in \mathbb{Z}_t^{t+T}}\E(k))$ is strongly connected; or that there exists a directed path between any two nodes.\\
(\textbf{A2}) $(A_i, B_i)$ is stabilizable,  $(A_i, C_i)$ is observable for all $i \in \mathbb{Z}^N$.\\
(\textbf{A3}) The states $x_i, i \in \mathbb{Z}^N$ are measurable.\\
(\textbf{A4}) $\U_i$ is a polytope and contains the origin in its interior for all $i \in \mathbb{Z}^N$.\\
Assumption (A1) is a standard necessary condition for time-varying network to achieve consensus \citep{ART:M05}.
(A2) is a reasonable requirement on individual agents although it is possible to relax it by introducing another output equation like $y_i^b(t)=E_i x_i(t)$ and requiring that $(A_i, B_i, E_i)$ be output stabilizable so that there exists a $F_i$ such that $(A+B_i F_i E_i)$ is Schur stable. This is, however, not done for the ease of presentation and to focus on the more novel aspects of the approach. (A3) is needed for the implementation of the Reference Governor, it is also necessary in the sense that without which, compliance of general constraint (\ref{eqn:xiui}) cannot be enforced. The assumption of (A4) is a reasonable expectation of $\U_i$ and is made to facilitate computational requirement.

\subsection{Internal Model Principle}

The approach of achieving output consensus using IMP for MAS where agents are constraint-free is now reviewed \citep{ART:FW76,BOO:KIF93,ART:WSA11}. The basic idea is to obtain, for every $i \in \mathbb{Z}^N$, a reference output from a reference model  having a diffusive term. Specifically,
\begin{subequations} \label{eqn:generatorsystem}
\begin{align}
\omega_i(t+1)&=S (\omega_i(t) + \sum_{j=1}^N a_{ij}(t) (\omega_j(t) - \omega_i(t))),\label{eqn:omegaitplus1}\\
y_i^r(t)&=Q \omega_i(t), \label{eqn:yr}
\end{align}
\end{subequations}
where $\o_i \in \mathbb{R}^{n_w}, S \in \R^{n_{\o} \times n_{\o}}$ is the reference model and $Q \in \mathbb{R}^{q \times n_\o}$ is common output among all agents. The output $y^r_i(t)$ generated using (\ref{eqn:yr}) is to be tracked by the output $y_i(t)$ of the $i^{th}$ system of (\ref{eqn:system}).  The value of $a_{ij}(t)$ changes with time according to $G(t)$. If $S$ is Schur stable, then $\omega_i$ converges to $0$ for all $i$ under (A1) \citep{ART:SS09}, a trivial case. Hence, it is typical to assume that \\
(\textbf{A5}) $(S,Q)$ is observable. \\
(\textbf{A6a}) $S$ has eigenvalues on the unit circle, or (\textbf{A6b}) $S$ has eigenvalues outside the unit circle.\\
When $S$ satisfies (A6a), $\omega_i$ of (\ref{eqn:omegaitplus1}) reaches consensus for all $i \in \mathbb{Z}^N$ under assumption (A1) \citep{ART:SS09}. When they do, so does $y_i^r$ for all $i \in \mathbb{Z}^N$.  If $y_i$ of (\ref{eqn:yi}) tracks $y_i^r$ asymptotically for every $i \in \mathbb{Z}^N$, $y_i$ reaches consensus as well for all $i \in \mathbb{Z}^N$. The tracking of $y_i$ to $y_i^r$ is made possible under IMP for a single system. This result is stated below as Lemma \ref{lem:IMP} for easy reference and notational setup. The proof is omitted as it is well known and can be found in references mentioned above. The result does require an additional assumption that ensures the existence of $L_i$, $\Pi_i$ and $\Gamma_i$ needed in the statement of the IMP.\\
(\textbf{A7})
$\left(
       \begin{array}{cc}
       A_i - \lambda I & B_i \\
       C_i & 0 \\
       \end{array}
\right)$ is full row rank for all $i \in \mathbb{Z}^N$ and for every eigenvalue $\lambda$ of $S$.

\begin{lemma}\label{lem:IMP}
Suppose (A2)-(A5), (A7) and, either (A6a) or (A6b), hold and consider one single system, system $i$, of (\ref{eqn:xikplus1}) and (\ref{eqn:yi}) and a reference system
\begin{align}\label{eqn:S}
  \o_i(t+1) & = S \o_i(t), \quad y_i^r(t)=Q \o_i(t)
\end{align}The unconstrained system of (\ref{eqn:xikplus1}) and (\ref{eqn:yi}) with
\begin{align}
u_i(t)&=K_i x_i(t) + L_i \omega_i(t) \label{eqn:uit} \\
L_i&:=\Gamma_i - K_i\Pi_i \label{eqn:Li}\\
e_i(t)&:= y_i(t) - y_i^r(t) = C_i x_i(t) -Q \o_i(t)
\end{align}
where $A_i+B_iK_i$ is Schur, combined with (\ref{eqn:S})
has the properties of \\
(i) $x_i(t) \rightarrow \Pi_i \omega_i(t)$ exponentially and
(ii) $e_i(t) \rightarrow 0$ exponentially \\
if and only if there exist matrices $\Pi_i \in \R^{n_i \times n_\omega}, \Pi_i \neq 0, \Gamma_i \in \R^{p_i \times n_\omega}$ such that
\begin{subequations}\label{eqn:AiPii}
\begin{align}
A_i \Pi_i - \Pi_i S &= - B_i \G_i \label{eqn:AiPiia}\\
C_i \Pi_i &= Q \label{eqn:AiPiib}
\end{align}
\end{subequations}
\end{lemma}

It is important to note that the stated property of (i) and (ii) above are for a single system without constraints.
In the presence of constraint (\ref{eqn:xiui}), the properties of Lemma 1 may not hold. 

\section{Choice of $S$} \label{sec:Choice}

The implications of (A6a) and (A6b) are now discussed.  Note that (\ref{eqn:omegaitplus1}), when collected over all $i \in \mathbb{Z}^N$, can be rewritten as
\begin{align}
\hat{\pmb{\o}}(t+1)= (S \otimes \P(t)) \hat{\pmb{\o}}(t),  \label{eqn:wtplus1}
\end{align}
where $\hat{\pmb{\o}}(t)=[\o_1^1(t) \cdots \o_N^1(t) \: \o_1^2(t) \cdots \o_N^2(t) \cdots \o_1^{n_w}(t) \cdots \\ \o_N^{n_w}(t) ]' \in \mathbb{R}^{N n_w}$. When (A6a) holds, the work of \cite{ART:SS09} shows that $\{\omega_i^j(t), i \in \mathbb{Z}^N\}$ achieve consensus exponentially for each $j \in \mathbb{Z}^{n_w}$ under (A1). On the other hand, if (A6b) holds, (\ref{eqn:wtplus1}) will generally not reach consensus or reach consensus only under strong assumptions on $\P(t)$ \citep{ART:SS09}. Besides needing  strong assumptions,  there is another reason for not considering (A6b) in the presence of constraint. It arises as constraint of (\ref{eqn:xiui}) imposes addition requirement on the structure of $\Gamma_i$ in (\ref{eqn:AiPii}). To see this, consider a similarity transformation
\begin{align} \label{eqn:coordinatechange}
\left(
  \begin{array}{c}
    \tilde{x}_i \\
    \o_i \\
  \end{array}
\right)
=\left(
    \begin{array}{cc}
      I & -\Pi_i \\
      0 & I \\
    \end{array}
  \right)\left(
  \begin{array}{c}
    x_i \\
    \o_i \\
  \end{array}
\right)
\end{align}
so that $u_i$ of (\ref{eqn:uit}) is expressed as
\begin{align} \label{eqn:uitGamma}
u_i(t)&=K_i(x_i(t)-\Pi_i \o_i(t)) + \Gamma_i \o_i(t)\nonumber \\
&=K_i \tilde{x}_i(t) + \Gamma_i \o_i(t).
\end{align}
In this form, it is clear that  $u_i(t) \rightarrow \Gamma_i \o_i(t)$ when $x_i(t) \rightarrow \Pi_i \o_i(t)$  under property (i) of Lemma \ref{lem:IMP}. If $S$ satisfies (A6b), $\o_i(t)$ becomes unbounded with increasing $t$ and, hence, $u_i(t)$ is unbounded for any non-zero $\Gamma_i$.
Equivalently, $u_i(t)$ is bounded only if $\Gamma_i =0$. Setting $\Gamma_i=0$ in (\ref{eqn:AiPiia}) results in $A_i\Pi_i = \Pi_i S$. The solution of which is well-known \citep{BOO:HJ91}: $\Pi_i = 0$ if and only if $spec(A_i) \cap spec(S) = 0$.  Since a meaningful result corresponds to having a non-zero $\Pi_i$, this means that $spec(A_i) \cap spec(S) \neq 0$, or in words,  every agent $A_i$ must contain the unstable eigenvalues of $S$ and that is a very strong requirement.




In view of the above, this work focuses on $S$ satisfying (A6a). Under which, there are two cases to consider: $S$ is Lyapunov stable and $S$ is not Lyapunov stable.  The former case has been discussed in \cite{INP:OD19}. This work focuses on the latter case where $S$ is, without loss of generality, given as a Jordan block. High-order Jordan blocks face difficulties in satisfying the conditions of (\ref{eqn:AiPii}) by most physical systems as well as complex controller design.
For this purpose, this work considers \\
(\textbf{A6})
$S=\left(
    \begin{array}{cc}
      1 & h \\
      0 & 1 \\
    \end{array}
  \right)$ \\
for some $h >0$ corresponding to the sampling period of the discrete-time system, and \\
\textbf{(A8)} $A_i$ has at least one eigenvalue of 1, for all $i \in \mathbb{Z}^N$.\\

These two assumptions have wide applicability in many physical systems. For example, (A8) holds for a large and important class of agents: helicopters, drones, land vehicles, surface/underwater sea vehicles etc. All of them have $1$ as one of its eigenvalues corresponding to the rigid body motion of the system. The next two lemmas relate to the implications of (A6) and (A8).
The first shows the properties of system (\ref{eqn:omegaitplus1}) under (A6) that are stated next for easy reference. It is a special case of theorem 2 of \cite{ART:SS09}.
\begin{lemma} \label{lem:omegavalue}
Suppose (A1) is satisfied, the system of (\ref{eqn:omegaitplus1}) with $S$ satisfying (A6) has the following properties:
(i) For each $j \in \mathbb{Z}^{n_w}$, $\o_i^j(t)$ reaches consensus for all $i \in \mathbb{Z}^N$, (ii) The consensus value of $\o_i(t)$ is given by $\o_i^2(t)=\bar{\o}^2$, $\o_i^1(t)=\bar{\o}^1 + h t \bar{\o}^2$ for some constants $\bar{\o}^2$ and  $\bar{\o}^1$.
\end{lemma}

\textbf{Proof:} 
(i) Under (A6), all eigenvalues of $S$ are on the unit circle. It follows from the result of \cite{ART:SS09} that consensus can be reached exponentially under (A1). In addition, the consensus value is $S^t  \bar{\o}(0)$, for some value of  $\bar{\o}(0)$. Let $(\bar{\o}^1(0), \bar{\o}^2(0))=(\bar{\o}^1, \bar{\o}^2)$ and noting that $S^t=\left(
             \begin{array}{cc}
               1 & ht \\
               0 & 1 \\
             \end{array}
           \right)$, the stated results follows. $\square$

A relevant issue when $S$ is given by (A6) is the value of $\Gamma_i$ in (\ref{eqn:AiPii}). From property (ii) of Lemma \ref{lem:omegavalue}, it is clear that $\o_i^1(t)$ is unbounded as $t$ increases. With $u_i(t)= K_i \tilde{x}_i(t) + \Gamma_i \o_i(t)$ from (\ref{eqn:uitGamma}), a necessary condition for $u_i(t) \in \U_i$ is that $\Gamma_i^1=0$, where $\Gamma_i^1$ is the first column of $\Gamma_i$.  A pertinent question is  ``Will the  $\Gamma_i$ obtained from the solution of (\ref{eqn:AiPii}) have $\Gamma_i^1=0$?". The next lemma addresses this question.

\begin{lemma}\label{lem:jordan2}
Consider the $i^{th}$ single system under the settings of Lemma \ref{lem:IMP} with (A2), (A5)-(A8) holding.
Then (i) Equation (\ref{eqn:AiPii}) admits a solution $(\Pi_i, \G_i)$ with $\Pi_i \neq 0, \G_i^1=0$ when $y_i^r \in \mathbb{R}$;
(ii) there exists a non-zero $Q \in \mathbb{R}^{n_w \times n_w}$ with $(S,Q)$ be observable such that (\ref{eqn:AiPii}) admits a solution $(\Pi_i, \G_i)$ with $\Pi_i \neq 0, \G_i^1=0$  for the case where $y_i^r \in \mathbb{R}^{n_w}$.
\end{lemma}
\textbf{Proof:}
(i) $(\Rightarrow)$ Let $\Pi_i=[\Pi_i^1 \quad \Pi_i^2]$ and $\G_i=[\G_i^1 \quad \G_i^2]$. The conditions of (\ref{eqn:AiPii}) can be expressed as
$A_i[\Pi_i^1 \quad \Pi_i^2] - [\Pi_i^1 \quad \Pi_i^1h + \Pi_i^2]=-B[\G_i^1 \quad \G_i^2]$
and $C_i [\Pi_i^1 \quad \Pi_i^2]= [Q^1 \quad Q^2]$, or,
\begin{subequations} \label{eqn:AiminusI}
\begin{align}
\left[
  \begin{array}{c}
    A_i-I \\
    C_i \\
  \end{array}
\right]\Pi_i^1 &= \left[
  \begin{array}{c}
    -B_i \G_i^1 \\
    Q^1 \\
  \end{array}
\right] \label{eqn:AiminusIa}\\
\left[
  \begin{array}{cc}
    A_i-I & B_i \\
    C_i & 0 \\
  \end{array}
\right]\left[
  \begin{array}{c}
    \Pi_i^2 \\
    \G_i^2 \\
  \end{array}
\right]&= \left[
  \begin{array}{c}
    \Pi_i^1 h \\
    Q^2 \\
  \end{array}
\right]\label{eqn:AiminusIb}
\end{align}
\end{subequations}
When $A_i$ has at least one eigenvalue of $1$ with non-zero eigenvector $\xi$, $rank(A_i-I) \leq n_i-1$. Since $(A_i,C_i)$ is observable under (A2), it follows from the PBH condition that $rank\left(
                                                       \begin{array}{c}
                                                         A_i-I \\
                                                         C_i \\
                                                       \end{array}
                                                     \right)=n_i$.
These two rank conditions show that $C_i$ does not lie in the row space of $A_i-I$.
Equivalently, $C_i$ has a component that is in the nullspace of $(A_i-I)$. Hence, $C_i \xi \neq 0$. When $y_i^r \in \mathbb{R}$, $Q^1$ is a non-zero scalar since $rank\left(
                                                       \begin{array}{c}
                                                         S-I \\
                                                         Q \\
                                                       \end{array}
                                                     \right)=n_w$
under (A5) and (A6). Hence, it is possible to scale $\xi$ by a constant $\gamma$ so that $C_i \xi \gamma=Q^1$ for any non-zero $Q^1$. This means that the choice of $\Pi_i^1= \gamma \xi$ and $\G_i^1=0$ satisfy (\ref{eqn:AiminusIa}). With this choice of $\Pi_i^1$, the right hand side of (\ref{eqn:AiminusIb}) is known. Existence of $\Pi_i^2$ and $\G_i^2$ for (\ref{eqn:AiminusIb}) follows from the full row rank assumption of (A7).
(ii) When $y_i^r \in \mathbb{R}^{n_w}$,  let $\xi$ be the eigenvector of $A_i$ corresponding to eigenvalue of $1$. Choose $Q^1=C_i\xi$ ( $C_i \xi \neq 0$ by the same reason given in the proof of (i)) with $Q^2$ arbitrary. As $Q^1$ is non-zero, this means that columns of $\left(
                                                       \begin{array}{c}
                                                         S-I \\
                                                         (Q^1 \ Q^2)  \\
                                                       \end{array}
                                                     \right)$
are linearly independent and that $(S,Q)$ is observable.  Then $\Pi_i^1 = \xi$ and $\G_i^1=0$ satisfy (\ref{eqn:AiminusIa}). Existence of $\Pi_i^2$ and $\G_i^2$  of (\ref{eqn:AiminusIb}) is again satisfied under (A7).
$\square$
\\

\section{Main Results} \label{sec:main}
The discussion in this section pertains to a single system and, for notational convenience, the reference of the $i^{th}$ agent is dropped in the various sets and matrices.

\subsection{MCAI set for unstable S} \label{sec:MCAI}
Property (ii) of Lemma \ref{lem:omegavalue} shows that $\o^1(t)$ is unbounded under (\ref{eqn:omegaitplus1}) when $S$ satisfies (A6). This means that $x(t)$ is unbounded following property (i) of Lemma \ref{lem:IMP}.  To ensure that $u(t)\in \U$ for each $t$, $\o(0)$ and $x(0)$ has to be limited to some appropriate invariant set.
This set is the maximal constraint admissible invariant set (MCAI) for the combined system of one \textit{single} $i^{th}$ system of (\ref{eqn:xikplus1})-(\ref{eqn:xiui}) together  with the IMP controller of (\ref{eqn:uit}) and reference system of (\ref{eqn:S}). Collectively, they are expressed as (where reference of $i$ is dropped)
\begin{align}
\left[
  \begin{array}{c}
    x(t+1) \\
    \omega(t+1) \\
  \end{array}
\right]&=\left[
          \begin{array}{cc}
            A+B K & BL \\
            0 & S \\
          \end{array}
        \right]\left[
  \begin{array}{c}
    x(t) \\
    \omega(t) \\
  \end{array}
\right] \label{eqn:xtwtplus1b}\\
u(t)&=\left[
             \begin{array}{cc}
               K & L \\
             \end{array}
           \right]
\left[
                   \begin{array}{c}
                     x(t) \\
                     \omega(t) \\
                   \end{array}
                 \right] \in \U \: \forall t \label{eqn:etait}
\end{align}
The MCAI of this system, $\O_\infty$, is
\begin{align}
\O_\infty=&\{ (x(0), \omega(0)): Kx(t)+L\o(t) \in \U \: \forall t, \nonumber\\
    & (x(t),\omega(t)) \textrm{ given by } (\ref{eqn:xtwtplus1b}) \} \label{eqn:Oinf}
\end{align}
Clearly, (\ref{eqn:xtwtplus1b}) is an unstable system since $S$ is unstable. In general, MCAI sets for unstable systems do not exist as state trajectories go unbounded from almost all initial states. Under the IMP framework with the appropriate assumptions, a non-empty MCAI set exists and is computable. Using  (\ref{eqn:Li}), (\ref{eqn:coordinatechange})  and (\ref{eqn:uitGamma}),  (\ref{eqn:xtwtplus1b}) is expressed as
\begin{align}
\left[
  \begin{array}{c}
    \tilde{x}(t+1) \\
    \omega(t+1) \\
  \end{array}
\right]&=\left[
          \begin{array}{cc}
            A+B K & 0 \\
            0 & S \\
          \end{array}
        \right]\left[
  \begin{array}{c}
    \tilde{x}(t) \\
    \omega(t) \\
  \end{array}
\right] \label{eqn:tildex}
\end{align}
In addition, $\o^1(t)$ is unbounded ( property (ii) of Lemma \ref{lem:omegavalue}) and $\Gamma^1=0$ (property (ii) of Lemma \ref{lem:jordan2}), (\ref{eqn:xtwtplus1b}) - (\ref{eqn:etait}) can be simplified to
\begin{align}
\left[
  \begin{array}{c}
    \tilde{x}(t+1) \\
    \omega^2(t+1) \\
  \end{array}
\right]&=\left[
          \begin{array}{cc}
            A+B K & 0 \\
            0 & 1 \\
          \end{array}
        \right]\left[
  \begin{array}{c}
    \tilde{x}(t) \\
    \omega^2(t) \\
  \end{array}
\right] \label{eqn:tildex2}\\
u(t)&=\left[
             \begin{array}{cc}
               K & \G^2 \\
             \end{array}
           \right]
\left[
                   \begin{array}{c}
                     \tilde{x}(t) \\
                     \omega^2(t) \\
                   \end{array}
                 \right] \in \U \: \forall t \label{eqn:uit2}
\end{align}
where the unbounded state of $\o^1(t)$ is omitted. Note that (\ref{eqn:tildex2}) is a Lyapunov stable system which, together with assumptions (A2)-(A4) and \\
\textbf{(A9)} System (\ref{eqn:tildex2})-(\ref{eqn:uit2}) is observable\\ 
satisfies all the conditions \citep{ART:GT91} needed for the existence of a non-empty MCAI set. With (A4), this MCAI set is a polytope, contains the origin in its interior, computable via an iterative procedure that terminates in a finite number of steps (Theorem 5.1 of \cite{ART:GT91} see also Remark \ref{rem:Oinf})  and has an expression of
\begin{align}\label{eqn:tildeOinf}
\tilde{\O}_\infty=\{(\tilde{x},\omega^2) \in \mathbb{R}^{n}\times \mathbb{R} : \tilde{H}_x \tilde{x} + \tilde{H}_w \omega^2 \le 1\}
\end{align}
for some matrices $\tilde{H}_x \in \mathbb{R}^{\ell \times n}, \tilde{H}_w \in \mathbb{R}^{\ell}$.
For technical reasons (see section \ref{sec:main}), a $\delta$-tightened $\tilde{\O}_\infty$ is also needed with $\delta > 0$ and is defined by
\begin{align}
	\tilde{\O}_\infty^\delta:&=\{(\tilde{x},\omega^2): \tilde{H}_x \tilde{x} + \tilde{H}_w \omega^2 \le 1-\delta\} \label{eqn:tildeOdelta}
\end{align}
Note that for a given $(\tilde{x},\o^2)$ the choice of $\o^1$ is not uniquely determined in (\ref{eqn:tildeOinf}) since it is satisfied by any $\o^1$ and $x$ such that $\tilde{x}=x - \Pi^1 \o^1 - \Pi^2 \o^2$. Of course, $\o^1$ is unique if $x$ is known.
Hence, when $(\tilde{x}, \o)$ is transformed back to $(x, \o)$, (\ref{eqn:tildeOinf}) becomes
\begin{align}
	& \O_\infty = \{ (x, \o) \in \mathbb{R}^{n}\times \mathbb{R}^{n_w} : H_x x + H_w \o \le 1\}\label{eqn:Oinf2}\\
	&\textrm{where } H_x :=\tilde{H}_x, H_w :=[ -\tilde{H}_x \Pi^1 \quad \tilde{H}_w -\tilde{H}_x \Pi^2]. \label{eqn:Hx}
\end{align}
Several properties related to this set are now given in the next lemma.
\begin{lemma} \label{lem:Oinf}
	Assume (A2)-(A9) hold for a single system.  The system of (\ref{eqn:xtwtplus1b})-(\ref{eqn:etait}) has the following properties: (i) $\tilde{\O}_\infty$ exists, is a full-dimensional set in $(\tilde{x}, \o^2)$ space and contains the origin; $\O_\infty$ is non-empty and is contained in a $k$-dimensional manifold with $k=n+n_w-1$; (ii) $(x, \o) \in \O_\infty$ implies $(\tilde{x}, \o^2) \in \tilde{\O}_\infty$ where $\tilde{x}=x - \Pi \o$ following (\ref{eqn:coordinatechange}); (iii) $(x(t),\o(t)) \in \O_\infty$ implies  $(x(t+1),\o(t+1)) \in \O_\infty$ where $(x(t+1),\o(t+1))$ follows (\ref{eqn:xtwtplus1b}) and that $u(t) \in \U$ for all $t$; (iv) $\omega(t)=S^t \omega(0)$ and $lim_{t \rightarrow \infty} x(t) = \Pi S^t \omega(0)$ for $(x(0), \omega(0)) \in \O_\infty$; (v) Suppose $(x(t),\omega(t))=(\Pi \eta, \eta) \in \O_\infty$ for some $\eta$ and some $t$, then $(x(t+1),\omega(t+1))=(\Pi S \eta, S \eta)$;
	(v) $\O_\infty$ is an unbounded subset of $\mathbb{R}^{n + n_w}$.
\end{lemma}

\textbf{Proof:} (i) The existence, full-dimensional and origin-inclusion property of $\tilde{\O}_\infty$ follows from the results given by Theorem 5.1 of \cite{ART:GT91} under assumptions (A2)-(A4) and (A9). Since $\O_\infty$ is obtained from $\tilde{\O}_\infty$ under (\ref{eqn:coordinatechange}), it is non-empty and lies in a subspace of dimension $n+n_w-1$.
(ii) Let $(x,\o) \in \O_\infty$. Using (\ref{eqn:Hx}) and (\ref{eqn:coordinatechange}) in (\ref{eqn:tildeOinf}) leads to
$\tilde{H}_xx+[-\tilde{H}_x \Pi^1 \quad -\tilde{H}_x \Pi^2 + \tilde{H}_w] \o \leq 1$, or rearranging,
$\tilde{H}_x \tilde{x} + \tilde{H}_w \o^2 \leq 1$
which implies $(\tilde{x}, \o^2) \in \tilde{\O}_\infty$. (iii) Let $(x, \o) \in \O_\infty$.  It follows from (ii) that  $(\tilde{x}, \o^2) \in \tilde{\O}_\infty$.
Since $\tilde{\O}_\infty$ is a MCAI set of (\ref{eqn:tildex2})-(\ref{eqn:uit2}),
$\tilde{H}_x(A+BK)\tilde{x}+\tilde{H}_w\o^2 \leq 1$
which implies, using (\ref{eqn:Hx}), (\ref{eqn:AiPiia}), (\ref{eqn:Li}) and (A6), that
$H_x((A+BK)x+BL\o)+H_w S\o\leq 1$
leading to $(x^+,\o^+))\in \O_\infty$ under the dynamics of (\ref{eqn:xtwtplus1b}).  Also the expression of $u(t)$ of (\ref{eqn:etait}) is equivalent to $u(t)$ of (\ref{eqn:uit2}) under $\Gamma^1=0$ and (\ref{eqn:Li}). Hence, $u(t) \in \U$ for all $t$.
(iv) Expression of $\omega(t)$ follows directly from the second equation of (\ref{eqn:xtwtplus1b}) and the value of $lim_{t \rightarrow \infty} x(t)$ is a consequence of (i) of Lemma \ref{lem:IMP}.
(v) Property (iv) states that $x(t) \rightarrow \Pi \o(t)$. Since $S$ satisfies (A6), $\o^1(t)=\o^1(0)+ \o^2(0)ht$. Hence, $x(t) = \tilde{x}(t)+\Pi^1 \o^1(t) + \Pi^2 \o^2(t) \rightarrow \Pi^2 \o^2(0)+ \Pi^1 (\o^1(0)+ ht \o^2(0))$ as $t \rightarrow \infty$. This shows that $\O_\infty$ is unbounded.
$\Box$

Four other sets associated with the $\O_\infty$ set are needed: the set of feasible $\omega$ at a given $x$, the set of feasible $x$ for a given $\o$, the set of admissible $\o$ and the set of admissible $x$ given by
\begin{align}
W(x):&=\{\omega: H_x x + H_w \omega \le 1\} \label{eqn:Wix}\\
\X(\omega):&=\{x: H_x x \le 1 - H_w \omega \} \label{eqn:Xio} \\
W_\infty:&=\{ \omega \in \mathbb{R}^{n_w}: \exists x \textrm{ such that } (x, \omega) \in \O_\infty \} \label{eqn:tWi}\\
\X_\infty:&=\{ x \in \mathbb{R}^{n}:  \exists \omega \textrm{ such that } (x, \omega) \in \O_\infty\} \label{eqn:Xinf}
\end{align}
As stated above,  $W_\infty$ and $\X_\infty$  are orthogonal projections of $\O_\infty$ onto the $\o$ and $x$ spaces respectively. The set $W_\infty$ is needed to construct admissible $u(t)$ and a more concrete characterization is needed.  Using property (i) of lemma \ref{lem:IMP} in the expression of (\ref{eqn:tildeOinf})
yields $W_\infty =\{(\o^1,\o^2) \in \mathbb{R}^2: \tilde{H}_\o \o^2 \leq 1\}$ or
\begin{align}
W_\infty:&=\{(\o^1,\o^2) \in \mathbb{R}^2:  \underline{\o^2} \leq \o^2 \leq \overline{\o^2} \} \label{eqn:Wiinfty}
\end{align}
for some lower and upper bounds, $\underline{\o^2}$ and $\overline{\o^2}$ respectively.
With the characterizations of $W_\infty$ and $\X_\infty$, a few observations are given.

\begin{remark} \label{rem:Xinf}
As a definition, (\ref{eqn:Xinf}) does not admit a clear geometrical interpretation of $\X_\infty$. One that does is to consider a different coordinate transformation by letting $\hat{x}=x - \Pi^1 \o^1$. The expression of (\ref{eqn:tildeOinf}) can be rewritten as $\tilde{H}_x (x - \Pi^1 \o^1) + (\tilde{H}_w - \tilde{H}_x \Pi^2) \o^2 \leq 1$, or,
$\hat{H}_x \hat{x} + \hat{H}_w \o^2 \leq 1$
This, together with (\ref{eqn:Wiinfty}), forms a combined system of
$\left(
  \begin{array}{cc}
    \hat{H}_x & \hat{H}_w  \\
    0 & 1 \\
    0 & -1 \\
  \end{array}
\right)
\left(
  \begin{array}{c}
    \hat{x} \\
    \o^2 \\
  \end{array}
\right)
\leq \left(
               \begin{array}{c}
                 1 \\
                 \overline{\o^2} \\
                 - \underline{\o^2} \\
               \end{array}
             \right)$.
Let the orthogonal projection of the set above onto the space of $\hat{x}$ be $G_\infty:=\{\hat{x} : \hat{G}_x \hat{x} \leq 1\}$. When converted back to the original $x$ space, one gets $\X_\infty=\{ x : \hat{G}_x (x - \Pi^1 \o^1) \leq 1, \o^1 \in \mathbb{R}\}$ which is equivalent to $\X_\infty = G_\infty \oplus \{span(\Pi^1)\}$ in set notation. An example of  $\X_\infty$ for a choice of $\Pi^1$ is depicted in Figure \ref{fig:Xinf}.
\begin{figure}[htbp!]
       \centering
       \includegraphics[scale=0.4]{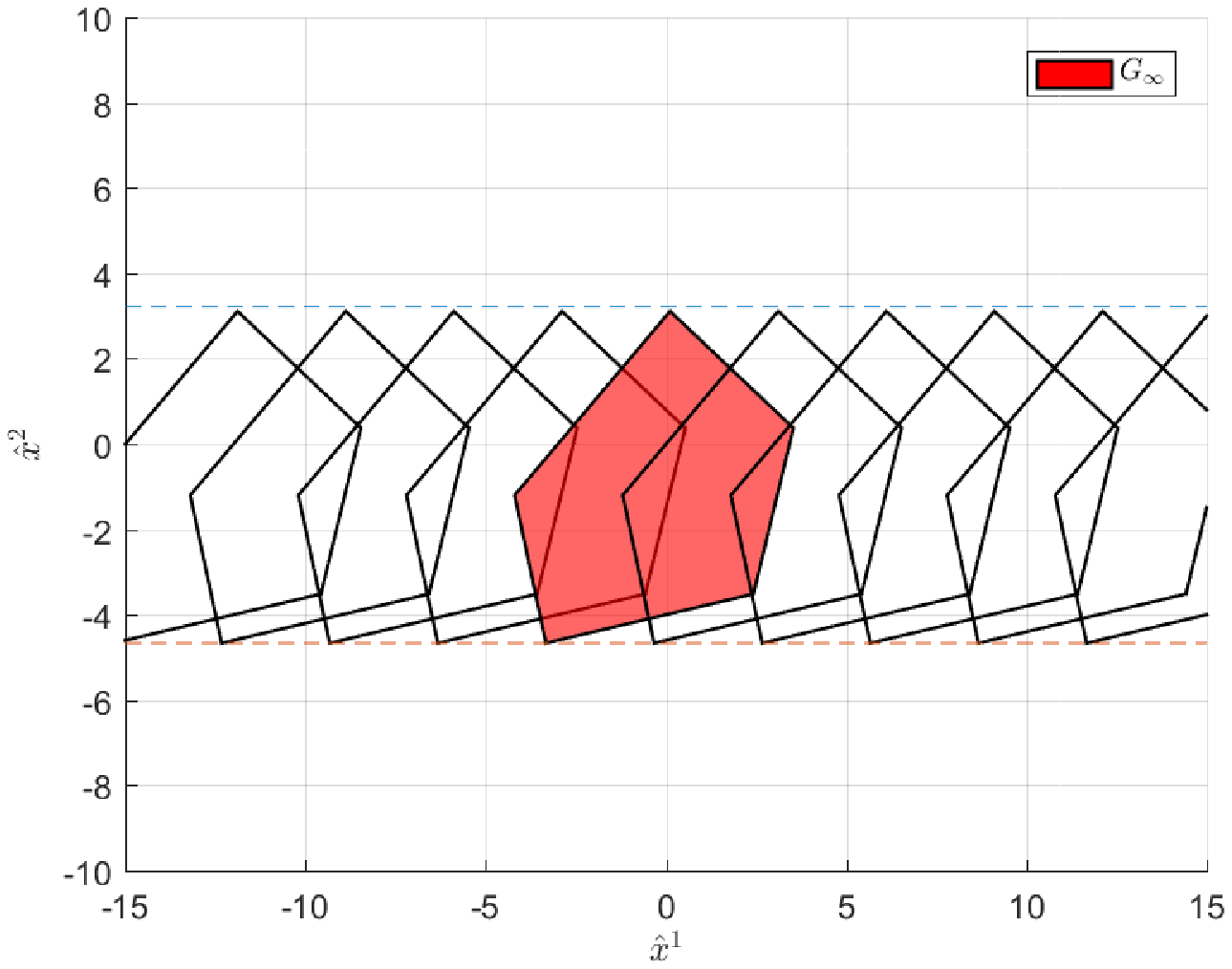}
       \caption{Depiction of $\X_\infty$ as $G_\infty \oplus \{span(\Pi^1)\}$ for a typical system with $\Pi^1 = [1\quad0]'$}\label{fig:Xinf}
\end{figure}
\end{remark}

\begin{remark}
The point $(0,\eta^2) \in \tilde{\O}_\infty$ is an equilibrium point of (\ref{eqn:tildex2}) and it corresponds to $(\Pi \eta, \eta)$ in the original $(x, \o)$ coordinate with $\eta=(\eta^1,\eta^2) \in W_\infty$.
This has the following interpretation:
For any $(x(0), \o(0)) \in \O_\infty$, the use of
$u(t)=Kx(t)+L\o(t)$ in (\ref{eqn:xikplus1}) drives $(x(t), \o(t))$ to the equilibrium point  $(\Pi \o(t), \o(t))$ of (\ref{eqn:tildex2}) with $u(t) \in \U$ for all $t$.
\end{remark}

\begin{remark} \label{rem:maxw2}
The constraints on $\o^2$ given by (\ref{eqn:Wiinfty}) define the maximum time rate of change of $\o^1(t)$ under (A6) for a given $\U$. Since $y^r(t)=Q\o(t)$, any $y^r(t)$ that has a time rate of change greater than $\overline{\o}^2$ or less than $\underline{\o}^2$ is untractable by $u(t) \in \U$ using the IMP approach.
\end{remark}

\begin{remark} \label{rem:Oinf}
Since $S$ contains eigenvalue of $1$, the computation of $\tilde{\O}_\infty$ may require many iterations to reach termination.  This problem and steps to minimize its effect have been discussed in Section V of \cite{ART:GT91} and \cite{ART:GO11}.  Specifically, a slightly truncated constraint
\begin{align}
W_\e&=\{(\o^1,\o^2) \in \mathbb{R}^2: \tilde{H}_\o \o^2 \leq 1 - \e\}\nonumber \\
&=\{(\o^1,\o^2) \in \mathbb{R}^2: \underline{\o^2}+\e \leq \o^2 \leq \overline{\o^2}-\e \} \label{eqn:Wepsilon}
\end{align}
for some $\e >0$ is used. This truncated set is used in all results henceforth.
\end{remark}

\subsection{The Reference Governor}\label{sec:RG}
\begin{figure}[htbp]
       \centering
       \includegraphics[scale=0.4]{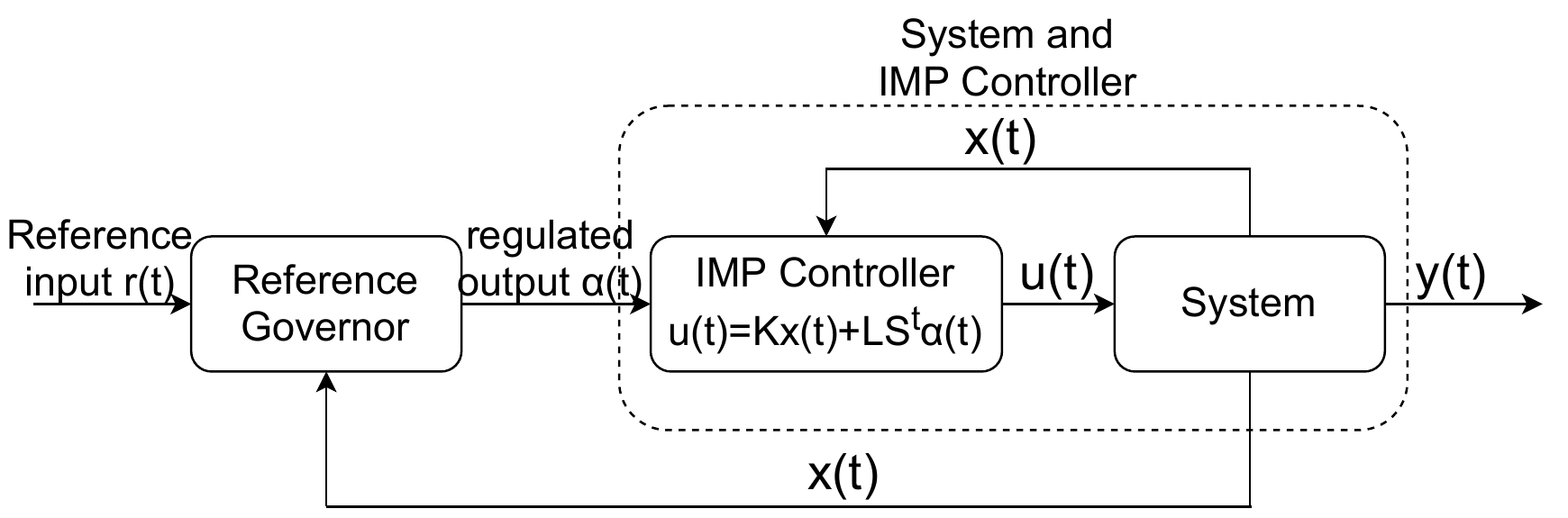}
       \caption{Plots of block diagram arrangement of the reference governor.}\label{fig:CG}
\end{figure}

Consider a single system of (\ref{eqn:xikplus1}) with the control input given by (\ref{eqn:uit}). Clearly, the value of $\o(t)$ of (\ref{eqn:uit}) can be seen as the reference input to this system.
Suppose output of this single system is to track a function $r(t) \in \mathbb{R}^{n_w}$,  the reference/command governor \citep{ART:CMP04,ART:GK02,ART:GO11} regulates the transition of the $\o(t)$ towards $r(t)$ so that constraint (\ref{eqn:xiui}) is satisfied at all time while keeping $\o(t)$ as close to $r(t)$ as possible. See Figure \ref{fig:CG} for a depiction of the block diagram arrangement of the command/reference governor within the system having IMP as the controller.

The reference governor (RG) used here defers from the conventional approach because of the presence of the unstable $S$. Suppose the affine function $r(t)=S^tr(0)$, the reference governor requires an additional state, $\a \in \mathbb{R}^{n_w}$ given by
\begin{gather}\label{eqn:alphatplus1}
\a(t)=\a(t-1)+\mu(t)(S^{-t}r(t)-a(t-1)), \a(0)=\a_0
\end{gather}
where $\a(0)=\a_0$ and $\a_0$ is obtained from
\begin{align}\label{eqn:alpha0}
	\a_0 (r(0), x(0) )&=arg min_\a \{ \|\a-r(0) \|_2^2: \nonumber\\
	&H_x x(0) + H_w \a \le 1, \a \in W_\e\}
\end{align}
The variable $\mu(t)$ is either zero or a function of $(x(t),r(t),\a(t-1))$. To state $\mu(t)$ precisely, consider the Linear Programming (LP) problem
\begin{subequations} \label{eqn:phistar}
	\begin{align}
		&\phi(t) :=\phi(x(t),r(t),\a(t-1))= \mathop{argmax}\limits_{\phi,\a_\phi \in \mathbb{R}^{n_w}} \: \phi \label{eqn:phi} \\
		&\a_\phi =\a(t-1) + \phi (S^{-t} r(t)-\a(t-1)) \label{eqn:alphi}\\
		&H_w S^t \a_\phi \le 1 - H_x x(t) \label{eqn:phiconstraint}\\
		&0 \le \phi \le 1 \label{eqn:boundonphi}
	\end{align}
\end{subequations}
In addition, noting that $\tilde{x}(t-1)=  x(t-1) - \Pi S^{t-1}\alpha(t-1)$ from (\ref{eqn:coordinatechange}), define
\begin{align}
	x^\dagger (t-1):=(A+BK)\tilde{x}(t-1)  \label{eqn:xdagger}
\end{align}
With the above definitions, the variable $\mu(t)$ of (\ref{eqn:alphatplus1}) is now given by
\begin{align}\label{eqn:mu}
	\mu(t)=\left\{
	\begin{array}{ll}
		\phi(t), & \hbox{if $(x^\dagger(t-1), \a^2(t-1)) \in \tilde{\O}_\infty^\delta$ ;} \\
		0, & \hbox{otherwise.}
	\end{array}
	\right.
\end{align}
where $\tilde{\O}_\infty^\delta$ is given by  (\ref{eqn:tildeOdelta}).
Note that $\phi$ of (\ref{eqn:phi}) defines a point, $\a_\phi$, on the line segment with endpoints  $\a(t-1)$ and $r(0)$. Maximizing $\phi$ in (\ref{eqn:phistar}) corresponds to $\a_\phi$ being farthest along the line segment towards $r(0)$ while remaining in $W(x(t))$.  Having obtained $\phi(t)$ from (\ref{eqn:phistar}), $\mu(t)$ is defined by (\ref{eqn:mu}) which, in turn, defines $\a(t)$ of (\ref{eqn:alphatplus1}). The control input to the single system is based on the IMP control law given by
\begin{align}\label{eqn:uiRG}
u(t)&=K x_i(t)+ L S^t \a(t)
\end{align}
The reference governor with IMP controller is now summarized in the following table assuming that $x(0)\in\X_\infty$. The properties of the command governor are given next.
\begin{algorithm}
	\caption{Reference Governor with IMP Controller}
	\label{alg:RGIMP}
	\begin{algorithmic}[1]\label{alg:A1}
		\STATE {Input: $x(t), r(t)$}
		\STATE {If $t=0$, compute $\a(0)$ from (\ref{eqn:alpha0}) and go to step (\ref{eqn:computeu}).}
		\STATE {Compute $x^\dagger(t-1)$ from $(x(t-1), \a(t-1))$ using (\ref{eqn:xdagger})}
		\IF {$(x^\dagger(t-1),\a^2(t-1)) \in \tilde{\O}^\delta_\infty$}
		\STATE {Compute solution of the Linear Program of (\ref{eqn:phistar}).}
		\STATE{Let $\mu(t)=\phi(t)$.}
		\ELSE
		\STATE {Let $\mu(t)=0$.}
		\ENDIF
		\STATE {Update $\a(t)$ as in (\ref{eqn:alphatplus1}).} \label{eqn:updateu}
		\STATE {Compute $u(t)$ as in (\ref{eqn:uiRG}) and apply to system.} \label{eqn:computeu}
	\end{algorithmic}
\end{algorithm}
\begin{theorem} \label{thm:1}
	Suppose (A2) - (A9) are satisfied and consider a single system of (\ref{eqn:xikplus1}) with  $x(0) \in \X_\infty$, $r(t)=S^t r(0)$, $r(0) \in W_\e$ is a
	given reference function.  Let $u(t)$ be given by (\ref{eqn:uiRG}) with $\a(t)$ from (\ref{eqn:alphatplus1}), $\mu(t)$ from (\ref{eqn:mu}) and $\a(0)$ from (\ref{eqn:alpha0}).  Then (i) $\a(t)$ exists and $u(t) \in \U$ for all $t \in \mathbb{Z}^+$; (ii) there exists a $t_f \in \mathbb{Z}^+$ such that $\a(t)=r(0)$  for all $t \ge t_f$ and $y(t)$ 
	approaches $y^r(t)$.
\end{theorem}
The proof of Theorem (\ref{thm:1}) is in the Appendix.

\section{The Multi-agent case}\label{sec:multi}
The case of multi-agent is now considered. All sets, vectors, matrices and other notations discussed in Section \ref{sec:main} that are for a single agent are now appended with an index corresponding to a specific agent in this section.

For consensus among the outputs of all the $N$ agents to be possible it is clear that $\o_i(t)$ needs to reach consensus for  all $i \in \mathbb{Z}^N$.  This, together with the fact that $\omega_i(t) \in W^i_\e$ for constraint admissibility, means that a necessary assumption for consensus is \\
\textbf{(A10)}: $\cap_i W^i_\e \neq  \emptyset$.\\
This assumption can be shown to hold under (A2)-(A4) and (A9). Specifically, $\tilde{\O}_\infty$ of (\ref{eqn:Oinf}) is a full dimensional set in $(\tilde{x},\o^2)$ space with the origin in its interior (see property (i) of lemma (\ref{lem:Oinf}). This means that $\tilde{x}=0, \o^2=0$ is in the interior of $\tilde{\O}_\infty$ which implies that $x=0, \o=0$ is inside $\O_\infty$.   Since $W_\e$ is the projection of $\O_\infty$ into the $\o$ space, the origin is contained in the interior of $W_\e$.  That this is stated as an assumption here is for easy reference and to emphasize the fact if (A10) is violated, consensus is not possible in the presence of constraint (\ref{eqn:xiui}).

The consensus of $\o_i(t)$ is now discussed.  In the form of (\ref{eqn:omegaitplus1}), $\omega_i(t)$ reaches consensus as given in Lemma \ref{lem:omegavalue}. However, the consensus value reached may not be feasible to the constraints of all agents. For constraint satisfaction and consensus, it is required that
\begin{align}
\lim_{t \rightarrow \infty} \o_i(t) = \bar{\o}(t) \textrm{ and } \bar{\o}(t) \in \cap_{i \in \mathbb{Z}^N} W^i_{\e}\quad \forall i \in \mathbb{Z}^N \label{eqn:limt}
\end{align}
where the non-emptiness of $\cap_{i \in \mathbb{Z}^N} W^i_{\e}$ is ensured by (A10).  To achieve (\ref{eqn:limt}), the system of (\ref{eqn:omegaitplus1}) is replaced by
\begin{align}
\o_i (t+1)&=\textrm{Proj}_{W^i_{\e}}(S(\omega_i(t)+\sum_{j \in N(i)} a_{ij}(t)(\o_j(t)-\o_i(t)))), \label{eqn:oitplus1b}
\end{align}
for all $i \in \mathbb{Z}^N$ where $\textrm{Proj}_W (\mu) := arg min_{\o} \{ \| \mu - \o \|^2 | \o \in W \}$. The consensus of $\o_i(t)$ having the property of (\ref{eqn:limt}) is now shown.
\begin{lemma} \label{lem:wconsensus}
  Suppose (A1), (A6) and (A10) are satisfied, $\o_i(t), i \in \mathbb{Z}^N$ of (\ref{eqn:oitplus1b}) reaches consensus satisfying (\ref{eqn:limt}).
\end{lemma}
\textbf{Proof :} Since $S$ under (A6) is of the form $S^t=\left(
        \begin{array}{cc}
          1 & ht \\
          0 & 1 \\
        \end{array}
      \right)
$, the second equation of (\ref{eqn:oitplus1b}) can be written as
\begin{align}\label{eqn:o2it}
\o^2_i (t+1)&=\mathop{\textrm{Proj}}\limits_{\Omega_i}((\o^2_i(t)+\sum_{j \in N(i)} a_{ij}(t)(\o^2_j(t)-\o^2_i(t))))
\end{align}
where $\Omega_i=\{\gamma \in \mathbb{R} : \underline{\o_i^2}+\e \leq \gamma \leq \overline{\o_i^2}-\e\}$ with $\underline{\o_i^2}, \: \overline{\o_i^2}$ being the limits on $\o^2$ given in (\ref{eqn:Wepsilon}) for agent $i$.  This is a constrained consensus problem which has been shown \citep{ART:NOP10,ART:LR14} to reach consensus asymptotically for all $i \in \mathbb{Z}^N$ with a consensus value of $\bar{\o}^2 \in \cap_{i} \Omega_i$ provided that the switching of the communication graph satisfies (A1) and that $\cap_{i} \Omega_i \neq \emptyset$ under (A10).

To show that $\o^1_i(t)$ also reaches consensus in (\ref{eqn:oitplus1b}), consider (\ref{eqn:oitplus1b}) for $t$ after $\o^2_i(t)$ has reached consensus. Using the transformation $z_i(t):=S^{-t} \o_i(t)$, it follows that
\begin{align}
&z_i(t+1)=S^{-(t+1)} \o_i (t+1) \nonumber\\
& =S^{-(t+1)} \mathop{\textrm{Proj}}\limits_{W^i_{\e}} (S(\omega_i(t)+\sum_{j \in N(i)} a_{ij}(t)(\o_j(t)-\o_i(t)))) \nonumber\\
  &  = S^{-(t+1)} \mathop{\textrm{Proj}}\limits_{W^i_{\e}} S^{t+1} (z_i(t)+\sum_{j \in N(i)} a_{ij}(t)(z_j(t)-z_i(t))) \nonumber\\
 & =  z_i(t) + \sum_{j \in N(i)} a_{ij}(t)(z_j(t)-z_i(t))) \label{eqn:zit}
\end{align}
where the last equation above follows from the fact that $\mathop{\textrm{Proj}}_{W^i_{\e}} \beta(t) = \beta(t)$ if $\beta(t) \in W^i_{\e}$. This last condition holds since $S$ and $W^i_\e$ have special structure under (A6) and (\ref{eqn:Wepsilon}) and $t$ is after the convergence of $\o^2_i(t)$ to $\bar{\o}^2$ with $\bar{\o}^2 \in \cap_{i} \Omega_i$. The expression of (\ref{eqn:zit}) is the standard consensus problem and $z_i(t)$ reaches consensus asymptotically when the switching communication graph satisfies (A1). When $z_i(t)$ reaches consensus, it follows from the transformation that $\o_i(t)$ also reaches consensus with (\ref{eqn:limt}) holding. $\square$

With this result, the overall algorithm for consensus reaching of (\ref{eqn:xikplus1}) is now stated.
\begin{algorithm}
	\caption{Multi-agent case}
	\label{Alg:MAS}
	\begin{algorithmic}[1]
		\STATE {Invoke the Reference Governor with IMP controller of Algorithm 1 with $(x_i(t), \o_i(t))$ as its input.}
		\STATE {Receive $\o_j(t)$ from all $j \in N_i(t)$. Compute $\o_i(t+1)$ using (\ref{eqn:oitplus1b}).}
		\STATE {Send $\o_i(t+1)$ to its neighbors and return to Step 1.}
	\end{algorithmic}
\end{algorithm}

The properties of Algorithm 2 are now stated.

\begin{theorem} \label{thm:2}
Suppose Assumptions (A1)-(A5), (A6), (A7)-(A10) hold and that $x_i(0) \in \tilde{\X}^i$ for all $i \in \mathbb{Z}^N$. System \ref{eqn:system} following Algorithm 1 above has the following properties for all $i \in \mathbb{Z}^N$ : (i) there exists a  $\bar{\omega}(t) \in  \cap_{i \in \mathbb{Z}^N} W^i_{\e}$ such that $\omega_i(t)$ approaches $\bar{\omega}(t)$ asymptotically and a $\bar{\o}^0 \in \cap_{i \in \mathbb{Z}^N} W^i_{\e}$ such that $\bar{\o}(t)=S^t \bar{\o}^0$. (ii) $u_i(t) \in \U_i$ for all $t$ ; (iii) there exists a finite $t_f$ such that $\a_i(t)= \bar{\omega}^0$ for all $t \ge t_f$ and $x_i(t) \rightarrow \Pi_i S^t\bar{\omega}^0$ (iv) $y^r_i(t)$ approaches $QS^t\bar{\omega}^0$ and $y_i(t)$ approaches $y_i^r(t)$ exponentially.
\end{theorem}

\textbf{Proof:}
(i) This result of $\bar{\omega}(t)$ reaches consensus follows from the result of Lemma \ref{lem:wconsensus}.
Since $\bar{\o}(t) \in \cap_{i \in \mathbb{Z}^N} W^i_{\e}$, it is easy to verify that $S^{-t} \bar{\o}(t) \in \cap_{i \in \mathbb{Z}^N} W^i_{\e}$ because of the special structure of $W^i_{\e}$ and $S$. In addition, $\bar{\o}(t+1)=S\bar{\o}(t)$ from Lemma \ref{lem:omegavalue} and the special structure of $S$ means that
$\bar{\o}^0 := S^{-t} \bar{\o}(t)$ is a constant. This establishes the second part of the property.
(ii) Step (\ref{eqn:computeu}) of Algorithm 1 ensures that $u_i(t)$ is the IMP control law obtained from (\ref{eqn:uiRG}) with $(x_i(t),S^t \a_i(t)) \in \O^i_\infty$ because of (\ref{eqn:phiconstraint}). Since $\O^i_\infty$ is constraint admissible, it follows from property (iii) of Lemma \ref{lem:Oinf} that $u_i(t)\in \U_i$ for all $t$.
(iii) When $\o_i(t)$ reaches consensus, it follows from property (i) above that $\bar{\o}(t+1)=S\bar{\o}(t)$. Consider $t$ after the time where $\o_i(t)$ has reached consensus, the setting of each individual system is exactly that given by Theorem \ref{thm:1} (with $r(t)$ of Theorem 1 replaced by $w_i(t)$). Hence, this property is the combination of properties (ii) and (iii) of Theorem \ref{thm:1}. (iv) When $x(t) \rightarrow \Pi_i S^t \bar{\o}^0$, it follows from Lemma \ref{lem:IMP} that $y^r_i(t) \rightarrow QS^t\bar{\omega}^0$ and $y_i(t) \rightarrow y_i^r(t)$ exponentially for all $i$.


\section{Enlarging the Domain of Attraction}\label{sec:global}
Theorem \ref{thm:2} applies to the case where $x_i(0)\in \X^i_\infty$. This can be restrictive for some systems especially when the `size' of $\X^i_\infty$ is small due to a high value of $K$. This section shows how this limitation can be circumvented for the common case where \\
(\textbf{A11}) $A_i$ is Lyapunov stable and has 1 as a simple eigenvalue. \\
The basic idea is to use a different control law when $x_i(t) \notin \X^i_\infty$ to drive $x_i(t)$ into $\X^i_\infty$. Once $x_i(t) \in \X_\infty^i$, the control law of steps (i) and (ii) of Algorithm I is then used.
\begin{lemma}\label{lem:RPi}
Assume (A11) holds and that $\O_\infty$ exists such that $\X_\infty$ is defined by (\ref{eqn:Xinf}). Then, $\mathcal{R}(\Pi^1) \subseteq \X_\infty$ where $\mathcal{R}(\Pi^1)$ is the Range space of $\Pi^1$.
\end{lemma}
\textbf{Proof:} Let $z \in \mathcal{R}(\Pi^1)$. This implies that there exists an $\alpha \in \mathbb{R}$ such that $z=\Pi^1 \alpha$. Choose $\o=[\a \quad 0]'$. It follows from (\ref{eqn:Oinf2}) that
$H_x (\Pi^1 \a) + H_w (\o)= H_x(\Pi^1 \a) - [\tilde{H}_x \Pi^1 \quad -\tilde{H}_x \Pi^2 + \tilde{H}_w]\left(
                                                                                                         \begin{array}{c}
                                                                                                           \a \\
                                                                                                           0 \\
                                                                                                         \end{array}
                                                                                                       \right)
 =0$ which implies that $(z, \o) \in \O_\infty$. Hence, $z \in \X_\infty$. $\square$

The design of the control law follows from the fact that $\Pi_i^1$ is the eigenvector of $A_i$ for the eigenvalue of $1$ from (\ref{eqn:AiminusIa}). Hence,  by expressing $A_i$ as $T \Lambda T^{-1}$ when $u_i(t)=0$, we have
\begin{align*}
  \lim_{t \rightarrow \infty} x_i(t) & = \lim_{t \rightarrow \infty} T \Lambda T^{-1} x_i(0) \\
  & = \lim_{t \rightarrow \infty} \sum_{j=1}^n q_j \lambda_j \mu_i^j(0)=\Pi_i^1 \mu_i^j(0)
\end{align*}
where $\mu_i(0)= T^{-1}x_i(0)$ and $q_j$ is the $j^{th}$ eigenvector of eigenvalue $\lambda_j$. The convergence of $x_i(t)$ above follows since $|\lambda_j|< 1$ for all $j$ except for $\lambda_j =1$ with the corresponding eigenvector $\Pi_i^1$.
This means that setting $u_i(t)=0$ leads to $x(t)$ approaches some point in $\mathcal{R}(\Pi_i^1) \subseteq \X^i_\infty$ where the last set inclusion follows from Lemma \ref{lem:RPi}.


With the above observation, the overall algorithm  is now given.
\begin{algorithm}
	\caption{Enlarging the Domain of $x_i(0)$}
	\label{alg:Enlarge}
	\begin{algorithmic}[1]
		\STATE {Given $x_i(0)$.}
		\IF {$x_i(t) \in \X_\infty^i$}
		\STATE {Invoke the reference governors of Algorithm 1.}
		\ELSE
		\STATE {let $u_i(t)=0$ and apply $u_i(t)$ as input to the $i^{th}$ system of (\ref{eqn:xikplus1})-(\ref{eqn:yi}).}
		\ENDIF
		\STATE {Receive $\o_j(t)$ from all $j \in N_i(t)$. Compute $\o_i(t+1)$ using (\ref{eqn:oitplus1b}).}
		\STATE {Send $\omega_i(t+1)$ to its neighbors and return to Step 2.}
	\end{algorithmic}
\end{algorithm}

\begin{remark}\label{eqn:computationofXi}
The set $\X^i_\infty$ appears in step (2) above for notational convenience. The characterization of $\X^i_\infty$ is not needed in the implementation of the Algorithm II. Specifically, $x_i(t) \in \X^i_\infty$ is replaced by the existence of $\rho$ such that $(x_i(t), \rho) \in \O_\infty^i$.
Of course, the testing of this condition can also be combined into step (2) of Algorithm \ref{alg:Enlarge} in the form of $min_g \{\|g-\o_i(t)\|^2 : g \in W_i(x_i(t))\}$.  Clearly, the quadratic program has a solution if and only if $x_i(t) \in \X_\infty^i$.
\end{remark}

\section{Numerical Examples}\label{sec:num}

\begin{figure}[htbp]
	\centering
	\subfigure[Plots of $y_i(t)$ versus $t$]{
		\begin{minipage}[t]{1\linewidth}
			\centering
			\includegraphics[width=2.5in]{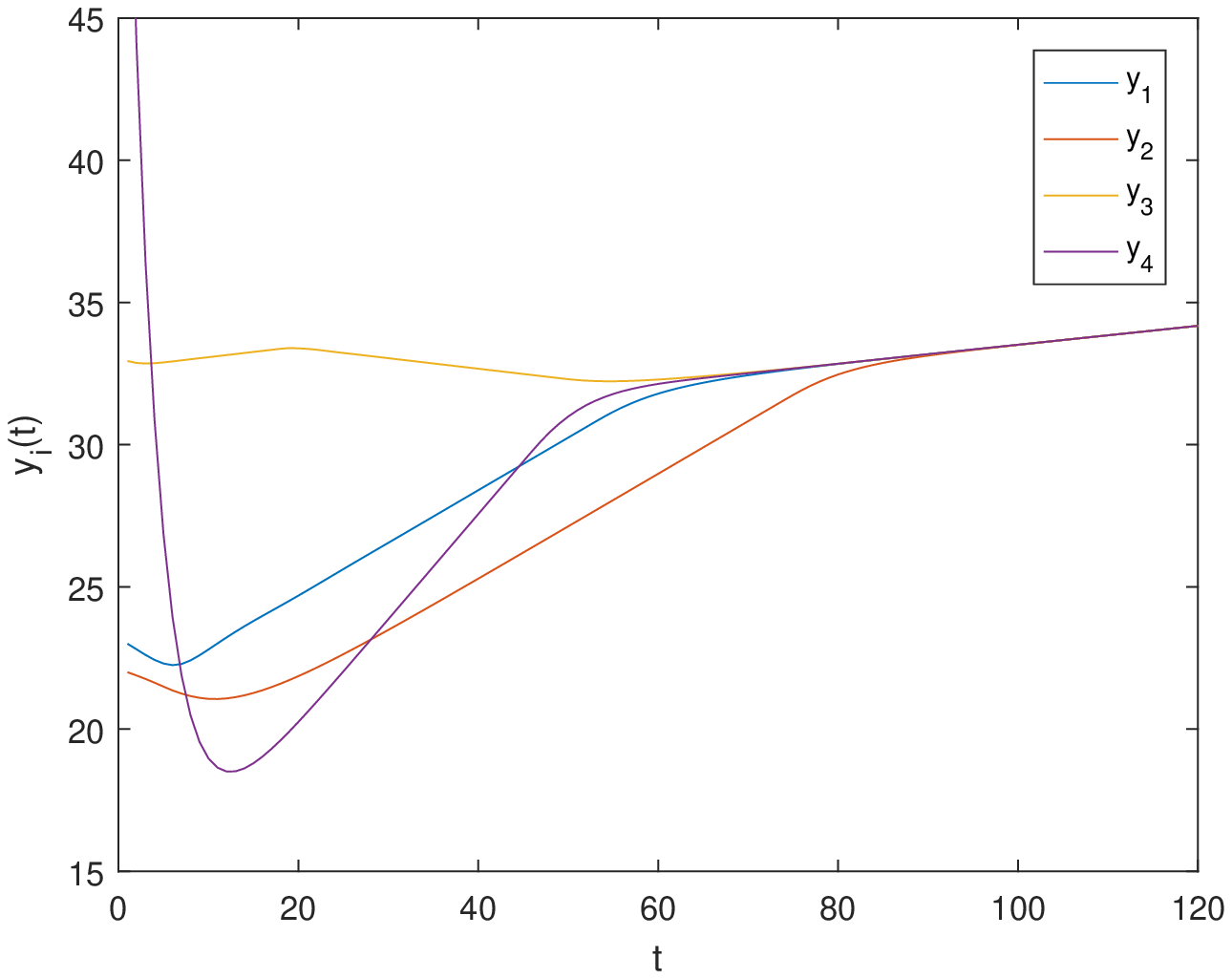}
		\end{minipage}
	}
	\subfigure[Plots of $y_i^r(t)$ versus $t$]{
		\begin{minipage}[t]{1\linewidth}
			\centering
			\includegraphics[width=2.5in]{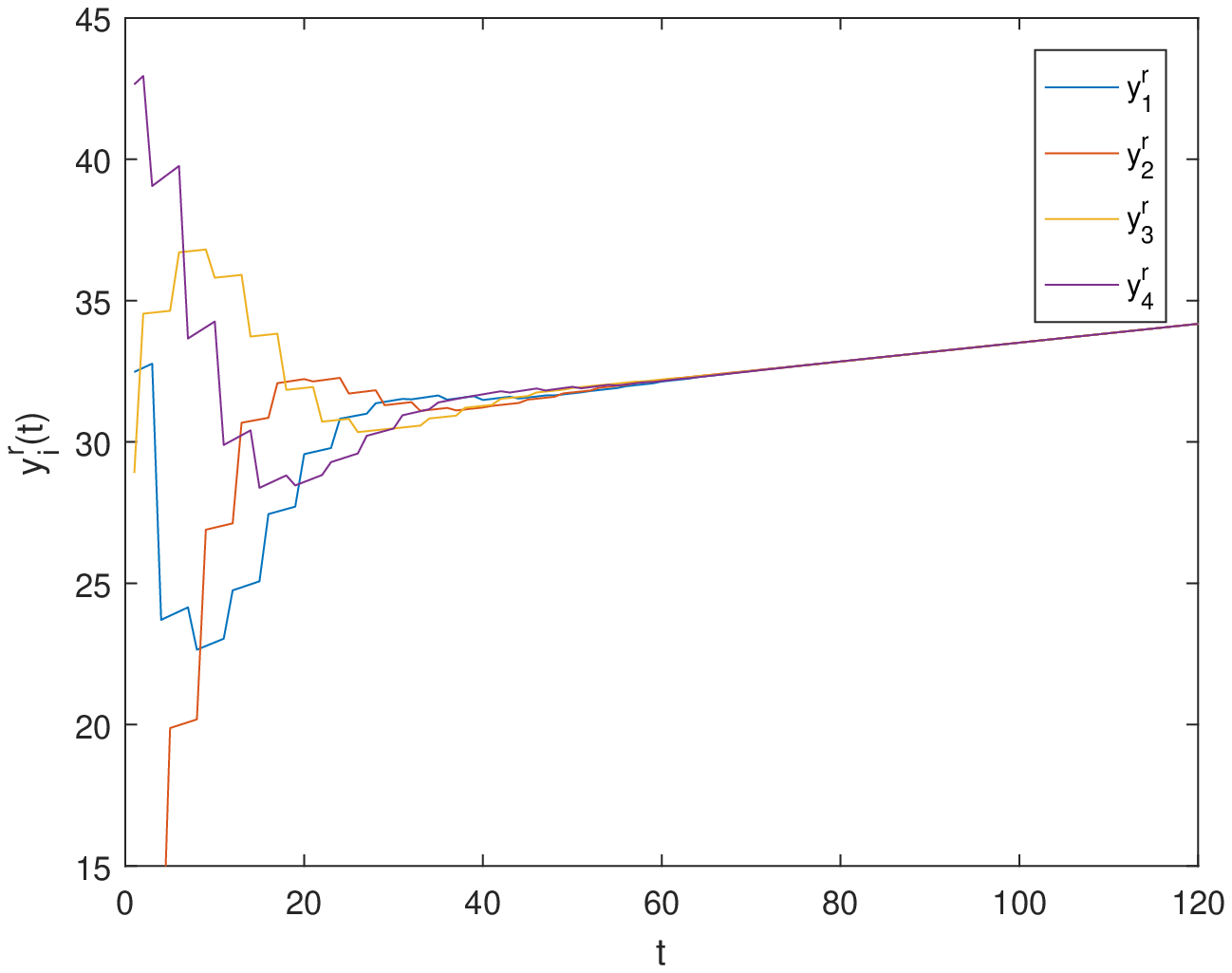}
		\end{minipage}
	}
	\caption{Graphs of $y_i(t)$ and $y_i^r(t)$ versus $t$}\label{Fig:OutAndRef6}
\end{figure}

\begin{figure}[htbp!]
	\centering
	\includegraphics[scale=0.45]{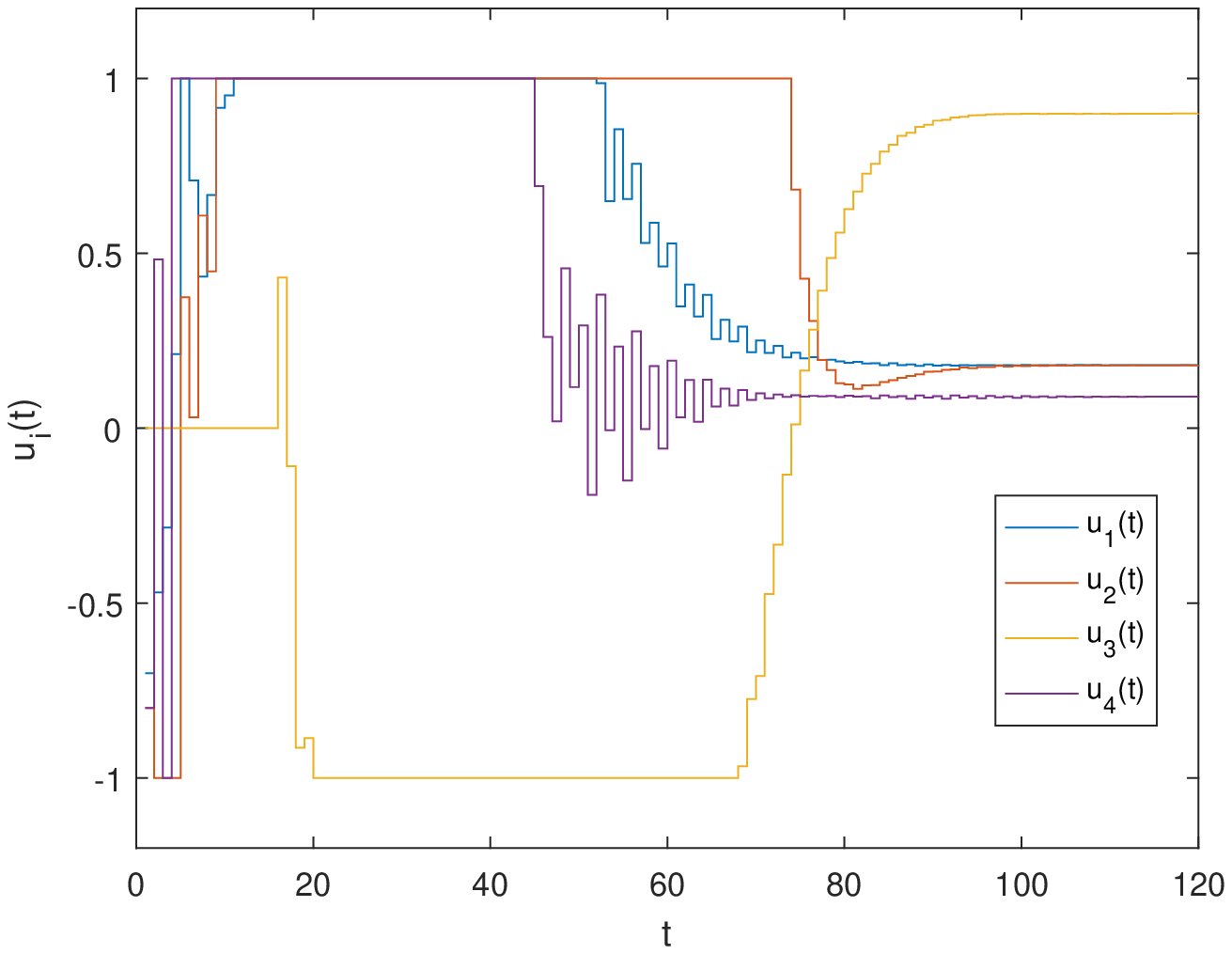}
	\caption{Plots of $u_i(t)$ versus $t$}\label{Fig:CON}
\end{figure}

The example given here is based on the continuous time example used in \cite{ART:WSA11}. It has $N = 4$, $p_i=1$ and $n_i = 3$ for all $i$ with the continuous-time agent converted to discrete-time with a Zero-Order-Hold sampler at a sampling period of $h = 0.37$. The graph $G(t)$ switches among $4$ graphs in the cyclic order of $G_1$ to $G_4$  where each $G_i$ has unit diagonal entries and zero otherwise except for $\P_{22}(G_1) = \P_{33}(G_2) = \P_{44}(G_3) = \P_{11}(G_4) = 0.6$, $\P_{23}(G_1) = \P_{34}(G_2) = \P_{41}(G_3) = \P_{12}(G_4) = 0.4$. Note that the union of $G_1$ to $G_4$ satisfies (A1).  The constraint set $\U_i = \{u_i: |u_i|<1\}$
The dynamics of the four systems are : $A_1 = \begin{bmatrix}
	1 & 0.3548 & 0.0594 \\
	0 & 0.8812 & 0.2954 \\
	0 & -0.5908 & 0.5858
\end{bmatrix}$, $A_2 = \begin{bmatrix}
	1 & 0.3538 & 0.0263 \\
	0 & 0.8946 & 0.09 \\
	0 & -0.3614 & -0.0089
\end{bmatrix}$, $A_3 = \begin{bmatrix}
	1 & 0.3036 & 0.0487 \\
	0 & 0.5134 & 0.2062 \\
	0 & -2.0623 & 0.1009
\end{bmatrix}$, $A_4 = \begin{bmatrix}
	1 & 0.363 & 0.0537 \\
	0 & 0.9463 & 0.2556 \\
	0 & -0.2556 & 0.4352
\end{bmatrix}$;
$B_1 = [0.0076\quad0.0594\quad0.2954]'$,
$B_2 = [0.0081\quad0.0527\quad0.1951]'$, $B_3 = [0.0066\quad0.0487\\ \quad0.2062]'$, $B_4 = [0.007\quad0.0537\quad0.2556]'$; $C_1 = C_2 = C_3 = C_4 = [1\quad0\quad0]$; $\Pi_1 = \Pi_2 = \Pi_3 = \Pi_4 = \begin{bmatrix}
	1 & 0 \\
	0 & 1 \\
	0 & 0
\end{bmatrix}$. Using them with $S$ of (A6), $\Gamma_1 = \Gamma_2 = [0\quad2]$, $\Gamma_3 = [0\quad10]$, $\Gamma_4 = [0\quad1]$,
$K_1 = [-2.3923\quad-4.99\quad-4.4074]$, $K_2 = [-3.8504\quad-8.9397\quad-2.9527]$, $K_3 = [-3.1011\quad0.9655\quad-3.8339]$, $K_4 = [-2.7975\quad-7.1542\quad-4.4122]$; $L_1 = [2.3923\quad6.999]$, $L_2 = [3.8504\quad10.9397]$, $L_3 = [3.1011\quad9.0345]$, $L_4 = [2.7975\quad8.1542]$; $Q = [1\quad0]$ for all $i$. Note that $A_i$ are Lyapunov stable with a simple eigenvalue at 1. In addition, $W^2_\infty=W^1_\infty=\{(w^1,w^2)\in\mathbb{R}^2:-0.5<w^2<0.5\}$, $W^3_\infty = \{(w^1,w^2)\in\mathbb{R}^2:-0.1<w^2<0.1\}$ and $W^4_\infty = \{(w^1,w^2)\in\mathbb{R}^2:-1<w^2<1\}$.\\
Two different sets of initial conditions are used:  the first set is such that $x_i(0)\in \X_\infty$ for all $i$ and the second has $x_i(0) \notin \X_\infty$ for some $i$. In the first case,  $x_1(0) = [23\quad-0.5\quad-0.2]'$, $x_2(0) = [22\quad-0.3\quad-0.1]'$,
$x_3(0) = [35\quad-0.3\quad0.22]'$, $x_4(0) = [54.5327\quad-33.0192\quad28.2356]'$
$w_1(0) = [32.4774\quad0.3968]'$, $w_2(0) = [9.4451\quad0.4]'$, $w_3(0) = [28.9\quad0.0793]'$, $w_4(0) = [42.6538\quad0.8]'$. The plots of $y_i(t)$ and $y^r_i(t)$ are shown in Figure \ref{Fig:OutAndRef6}(a) and \ref{Fig:OutAndRef6}(b) respectively. Clearly, Figure \ref{Fig:OutAndRef6}(a) shows that consensus is reached for $y_i(t)$ while \ref{Fig:OutAndRef6}(b) shows the same for $y^r_i(t)$. Notice also the rapid switches of $y^r_i(t)$ in Figure \ref{Fig:OutAndRef6}(b) for the initial values of $t$ resulting from the projection operations of $\o_i(t)$ into $W_\e^i$. The fact that $y^r_i(t)$ reaches consensus among all $i$ and follows the reference model is as predicted in property (i) of Theorem \ref{thm:2}.  The corresponding plots of $u_i(t)$ are depicted in Figure \ref{Fig:CON} and clearly show that $u_i(t) \in \U_i$ for all $i$ and all $t$.\kkk
\begin{figure}[htbp!]
	\centering
	\includegraphics[scale=0.45]{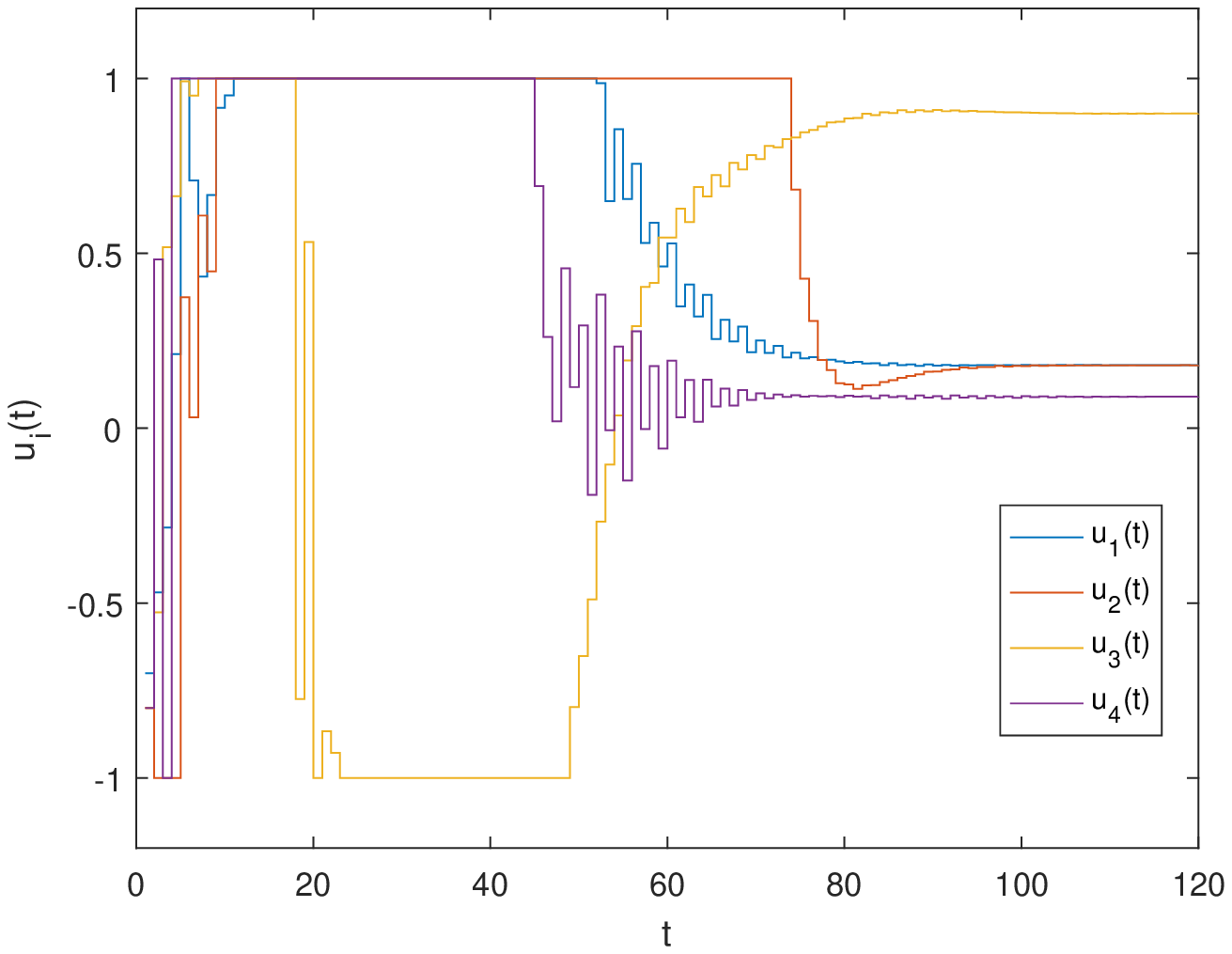}
	\caption{Plots of $u_i(t)$ versus $t$}\label{Fig:CON6}
\end{figure}

\begin{figure}[htbp]
	\centering
	\subfigure[Plots of $y_i(t)$ versus $t$]{
		\begin{minipage}[t]{1\linewidth}
			\centering
			\includegraphics[width=2.5in]{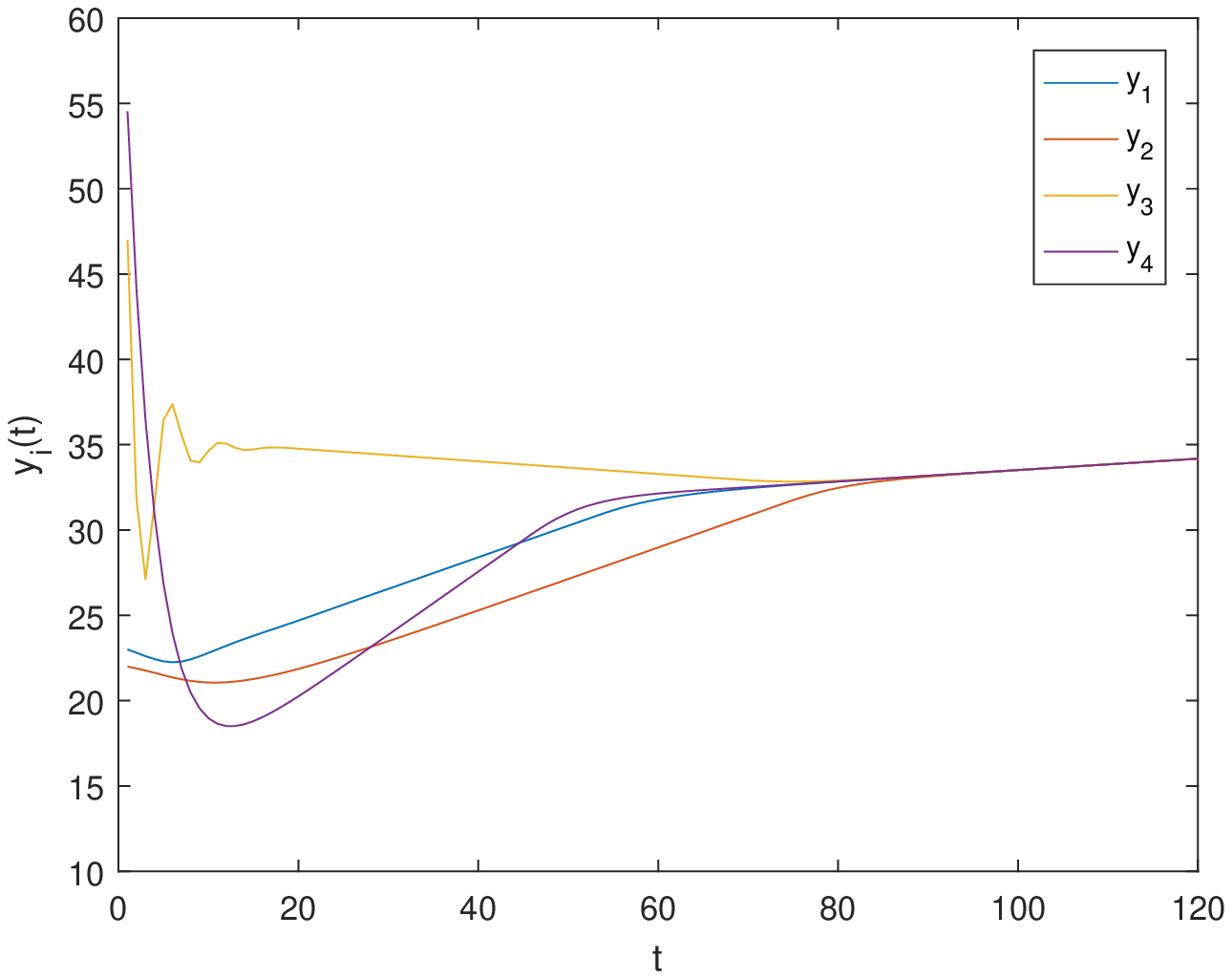}
		\end{minipage}
	}
	\subfigure[Plots of $y_i^r(t)$ versus $t$]{
		\begin{minipage}[t]{1\linewidth}
			\centering
			\includegraphics[width=2.5in]{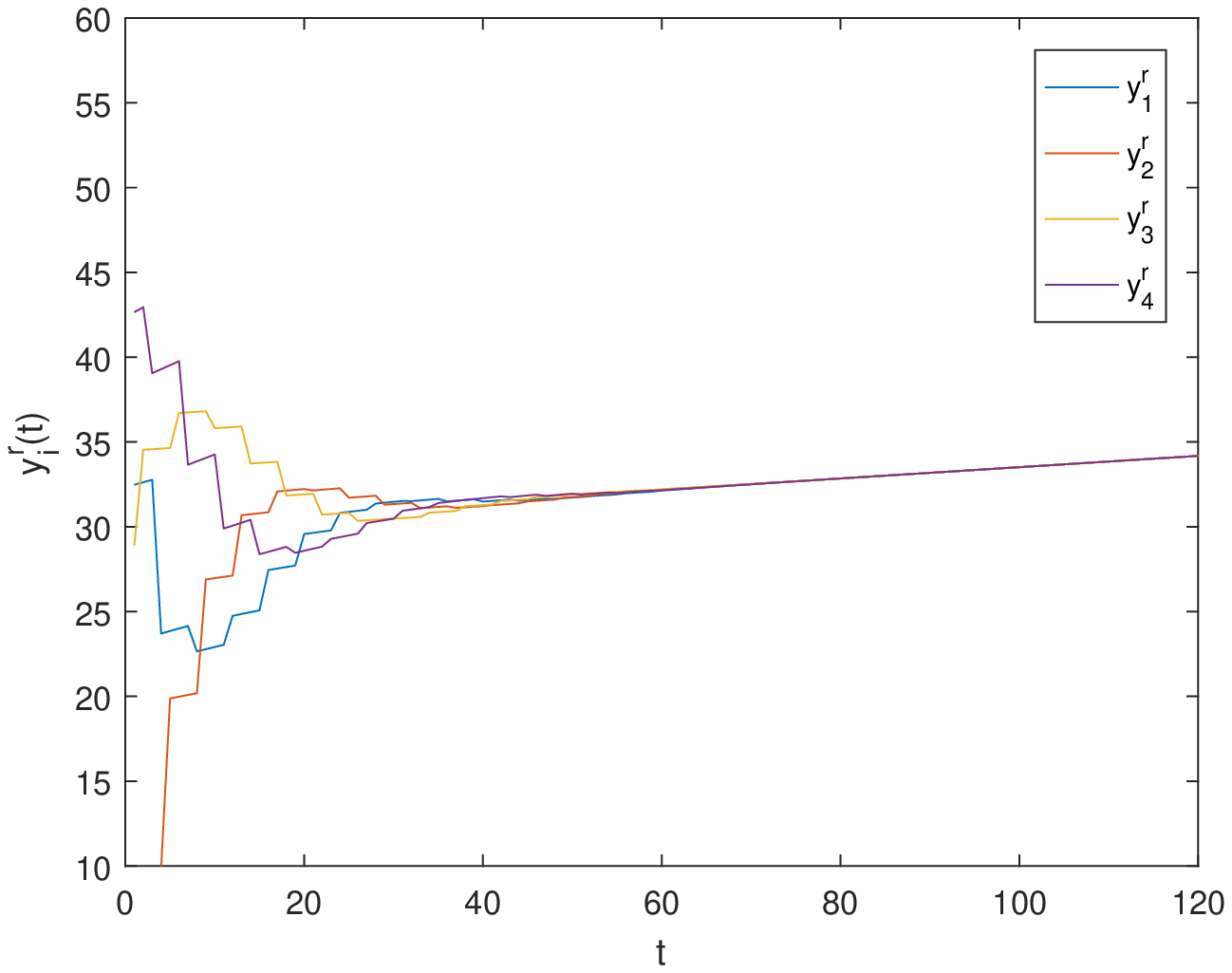}
		\end{minipage}
	}
	\caption{Graphs of $y_i(t)$ and $y_i^r(t)$ versus $t$}\label{Fig:OutAndRef}
\end{figure}

In the second case, all $x_i(0)$ are the same as the first except $x_3(0) = [47\quad-45\quad-32]'$.
Note that $x_3(0) \notin \X^3_\infty$ and the procedure of section \ref{sec:global} is used.
The plots of $y_i(t)$ and $y^r_i(t)$ are shown in Figure \ref{Fig:OutAndRef}(a) and \ref{Fig:OutAndRef}(b) respectively.
Besides the consensus results, of particular interest is the initial wavy motion of $y_3(t)$ for small values of $t$. It corresponds to the switching between two control laws as described in Step (ii) of Algorithm (\ref{alg:Enlarge}) since $x_3(0) \notin \X_\infty^3$. Also, in Figure \ref{Fig:CON}, $u_3(t)$ is equal to 0 for the first several iterations, corresponding to Algorithm (\ref{alg:Enlarge}).\kkk

\section{Conclusions}\label{sec:con}
This paper describes the procedure for achieving output consensus among heterogeneous agents where each agent is subject to constraints on its inputs in a multi-agent environment.  The communication among agents is described by a time-varying directed graph.  The approach is based on the Internal Model Principle where a specific unstable reference model is used and for a large common family of agents. One significant novelty of the approach is the characterization of the maximal constraint admissible invariant (MCAI) set for the unstable combined system.  Using this set, consensus and constraint admissibility is achieved via a two-step optimization process, the first is a projection step that brings the reference inputs to the common feasible input space of all agents, the second is a novel reference/command governor that ensures satisfaction of constraints of each agent. Both convergence to consensus and constraint admissibility are proved under reasonably mild assumptions.


\section{Appendix}

\textbf{Proof of Theorem \ref{thm:1}}\\
(i) The existence of $\a(t)$ depends on the existence of $\phi(t)$. We begin with $\a_0$. Since $x(0) \in \X_\infty$, the set of $\a$ such that $ H_x x(0) + H_w \a \le 1$ is non-empty. This, together with the fact that $\|\a-r(0) \|_2^2$ is continuous in $\a$ means that $\a_0$ exists. The state $\a(t)$ for all $t \ge 1$ also exists since $\phi(t)$ exists as the LP problem of (\ref{eqn:phistar}) has at least one feasible point of $\phi=0$. The fact that $u(t) \in \U$ follows from property (iii) of Lemma \ref{lem:Oinf} because $(x(t), S^t \a_\phi) \in \O_\infty$ as imposed by (\ref{eqn:phiconstraint}).

(ii) A few properties and notations are first established. We begin by noting that $\mu(t)$ is obtained from two expressions as given by (\ref{eqn:mu}) depending on the set inclusion condition of $(x^\dagger(t-1),\a^2(t-1)) \in \tilde{\O}_\infty^\delta$. To distinguish these two sequences of $\{\mu(t)\}$, let the time sequence be
$T=\{0,\cdots, t_0, t_0+1,\cdots, t_0+\bar{t}_0, t_1, t_1+1,\cdots,t_1+\bar{t}_1,t_2,\dots\}$
where $T_a:=\{t_0,t_1,t_2,\cdots,t_i,\cdots\}$ refers to set of time instants where $\mu(t)$ is given by the solution of the LP and 
$T_b$ is the remaining time instants such that $T=T_a \cup T_b$ and $T_a \cap T_b=\emptyset$.
Correspondingly, $\mu_a=\{\mu(t)|t\in T_a\}$ and $\mu_b=\{\mu(t)|t\in T_b\}$.
We first show that $T_a$ is an infinite subsequence of $T$. This is done by showing that $t$ cannot stay in $T_b$ forever.  This is shown in lemma \ref{lem:Tb1}.
The next property is
\begin{align}
	\hspace{-3mm}r(0)-\a(t)=(1-\mu(t))\cdots(1-\mu(0))(r(0)-\a(0)) \label{eqn:r0}
\end{align}
To show the above, note from (\ref{eqn:alphatplus1}) that
$\alpha(t)=(1-\mu(t))\alpha(t-1)+\mu(t)r(0)$
which implies that $r(0)-\alpha(t)=r(0)-\mu(t)r(0)-(1-\mu(t))\alpha(t-1)=(1-\mu(t))(r(0)-\alpha(t-1))$.
Repeated application of the above for decreasing value of $t$ yields (\ref{eqn:r0}). For notational convenience,
let $\xi(t):=(1-\mu(t))(1-\mu(t-1))\cdots(1-\mu(0))$. Several obvious properties of $\xi(t)$ are :
\textbf{P1:} $\xi(t)=(1-\mu(t))\xi(t-1)$;
\textbf{P2:} $0\leq \xi(t)\leq 1 \: \forall t \geq 0$;
\textbf{P3:} $\xi(t)\leq \xi(t-1)$ with $\xi(t)=\xi(t-1)$ when $t \in T_b$.
Following P3, $\xi(t)$ and $\a(t)$ change values only when $t \in T_a$.
Note also that $x(t_i)$ can be expressed as,
\vspace{-8mm}
\small
\begin{align}
	&x(t_i)=Ax(t_i-1)+B(Kx(t_i-1)+LS^{t_i-1}\alpha(t_i-1))\nonumber\\
	&=(A+BK)x(t_i-1)+BLS^{t_i-1}\alpha(t_i-1)\nonumber\\
	&=(A+BK)(\tilde{x}(t_i-1)+\Pi S^{t_i-1}\alpha(t_i-1))+BLS^{t_i-1}\alpha(t_i-1)\nonumber\\
	&=(A+BK)\tilde{x}(t_i-1)+((A+BK)\Pi+B(\Gamma-K\Pi))S^{t_i-1}\alpha(t_i-1)\nonumber\\
	&=(A+BK)\tilde{x}(t_i-1)+\Pi S^{t_i}\alpha(t_i-1) \label{eqn:xti}
\end{align}\normalsize
where $u(t_i-1)$ of (\ref{eqn:uiRG}), (\ref{eqn:coordinatechange}), (\ref{eqn:Li}) and (\ref{eqn:AiPiia}) are used.

The next fact is to note that for all $t \in T_a$, the solution, $\phi(t)$, of the LP is such that one of the constraints of (\ref{eqn:phiconstraint}) holds as an equality.  This is so because solution of LP occurs at the boundary of its constraint set. If $\phi(t)$ is such that (\ref{eqn:phi}) is an equality, then $\phi=1$ and the corresponding value of $\a_\phi$ of (\ref{eqn:alphi}) is $\a_\phi = S^{-t}r(t)=r(0)$. In which case, the output of Reference Governor will have already achieved its final value and it is easy to verify that $\a(t)=r(0)$  for all $t$ thereafter.
With these facts, property (ii) of Theorem 1 is now proved by contradiction. Suppose $\lim_{t \rightarrow \infty} \a(t) \neq r(0)$ and
consider the constraint (\ref{eqn:phiconstraint}) of the LP at some $t_i\in T_a$. Since the optimizer of LP occurs at the boundary of the feasible set,
some specific row, $k_i$, of (\ref{eqn:phiconstraint}) holds as an equality, or
$[H_xx(t_i)+H_wS^{t_i}\alpha(t_i)]_{k_i}=1$. Using (\ref{eqn:xti}) and (\ref{eqn:alphatplus1}) in this row yields
\begin{align}
	&[H_x(A+BK)\tilde{x}(t_i-1)+(H_x\Pi+H_w)S^{t_i}\alpha(t_i-1)\nonumber\\
	&+H_wS^{t_i}\mu(t_i)(r(0)-\alpha(t_i-1))]_{k_i}=1 \label{eqn:xtilde2}
\end{align}
\kkk
Using (\ref{eqn:r0}), P1 and
$r(0)-\alpha(t_i-1)=\xi(t_i-1)(r(0)-\alpha(0))$, the expression in the third term of (\ref{eqn:xtilde2}) is
$\mu(t_i)(r(0)-\alpha(t_i-1))=\mu(t_i)\xi(t_i-1)(r(0)-\alpha(0))=(\xi(t_i-1)-\xi(t_i))(r(0)-\alpha(0))$.
With the result above, (\ref{eqn:xtilde2}) becomes
\vspace{-5mm}
\small
\begin{align}
	&[H_wS^{t_i}(\xi(t_i-1)-\xi(t_i))(r(0)-\alpha(0))]_{k_i}=\nonumber\\
	&[1-(\tilde{H}_x(A+BK)\tilde{x}(t_i-1)+\tilde{H}_wS^{t_i}\alpha(t_i-1))]_{k_i}\label{eqn:xtilde4}
\end{align}
\normalsize
Since $t_i \in T_a$, it implies that $((A+BK)\tilde{x}(t_i-1),\alpha^2(t_i-1)) \in \tilde{\O}^\delta_\infty$, or that $\tilde{H}_x(A+BK)\tilde{x}(t_i-1)+\tilde{H}_wS^{t_i}\alpha(t_i-1)\leq 1-\delta$. Using this result in (\ref{eqn:xtilde4}), one gets $[H_wS^{t_i}(\xi(t_i-1)-\xi(t_i))(r(0)-\alpha(0))]_{k_i}\geq \delta>0$. Let
\begin{equation}\label{eqn:Pdelta}
	\delta_i:= [H_wS^{t_i}(\xi(t_i-1)-\xi(t_i))(r(0)-\alpha(0))]_{k_i}
\end{equation}
which means $\delta_i \geq \delta$. Note that $\xi(t_i-1)-\xi(t_i)$ is a scalar and $\xi(t_i-1)-\xi(t_i)\geq 0$ from P3, so $[H_wS^{t_i}(r(0)-\alpha(0))]_{k_i}$ must be a positive scalar from (\ref{eqn:Pdelta}). In addition,  $S^{t_i}=\begin{pmatrix}
	&1 &ht_i\\
	&0 &1
\end{pmatrix}$ and all other terms are constants, it can be verified that
$[H_wS^{t_i}(r(0)-\a(0))]_{k_i}= \beta_i^1 t_i + \beta_i^2$
for some constants  $\beta_i^1$ and $\beta_i^2$ with $\beta_i^1 >0$ and $\beta_i^2$ is a bounded scalar for $t_i$ that are sufficiently large.
These facts mean that (\ref{eqn:Pdelta}) can be written as
\begin{equation}\label{eqn:Pdiff2}
	\begin{aligned}
		\xi(t_{i-1})-\xi(t_i)&=\frac{\delta_i}{\beta_i^1 t_i+\beta_i^2}
	\end{aligned}
\end{equation}
Since $T_a$ is an infinite sequence, applying the above for $t_j$ for $j=i, i+1, \cdots$ and adding these equations up yields
$\xi(t_i)-\xi(\infty)= \sum_{t_i}^{\infty} \frac{\delta_i}{\beta_i^1 t_i+\beta_i^2}$.
Using Lemma \ref{lem:math} on the right hand side of the above, this leads to
$\xi(t_i)-\xi(\infty)= \infty. $
However, since $\xi(t_i)$ and $\xi(\infty)$ are both bounded (from P2), their difference can not be infinity. This leads to a contradiction to the assumption that  $\lim_{t \rightarrow \infty}\a(t) \neq r(0)$. Hence, there exists a finite $t_f$ such that $\a(t_f)=r(0)$. When $\a(t_f)=r(0)$, $u(t)=Kx(t)+LS^tr(0)$ for all $t \geq t_f$, then $y(t) \rightarrow y^r(t)$ exponentially following property (i) of Lemma \ref{lem:IMP}.

\begin{lemma}\label{lem:Tb1}
	Suppose the $(x(t),\a(t))$ is such that $t \in T_b$, then there exists a finite $\bar{t} \in \mathbb{Z}^+$, independent of $t$, such that 
	$(t+\bar{t}) \in T_a$.
\end{lemma}
\textbf{Proof:} Suppose the above is not true or that $t+1, t+2, \cdots, t+\infty$ are all in $T_b$. This means that $\mu(t+i)=0, \a(t+i)=\a(t)$ for all $i=1,2,\cdots$ and $(\tilde{x}(t),\a(t))$ follows the dynamics given by (\ref{eqn:tildex2}).
Since $(\tilde{x}(t),\alpha^2(t))\in \tilde{\mathcal{O}}_\infty$, this means that  $\tilde{H}_x\tilde{x}(t)+\tilde{H}_w\alpha^2(t)\leq 1$. Repeating this till $t+\bar{t}$,
it follows that $\tilde{H}_x\tilde{x}(t+\bar{t})+\tilde{H}_w\a^2(t+\bar{t})\leq 1$.
Expressing $\tilde{x}(t+\bar{t})$ in terms of $\tilde{x}(t)$ under (\ref{eqn:tildex2}), the above becomes
$\tilde{H}_x(A+BK)^{\bar{t}}\tilde{x}(t)+ \tilde{H}_w \a^2(t+\bar{t}) \leq 1.$
Since $(A+BK)$ is stable, choose $\bar{t}$ such that $\tilde{H}_x(A+BK)^{\bar{t}}\tilde{x}(t)\leq \epsilon-\delta$. As $\tilde{x}(t)$ is contained in a bounded set for all $t$ as seen from $\tilde{\O}_\infty$ of (\ref{eqn:tildeOinf}), $\bar{t}$ is an upper bound that is independent of $t$. This, together with the fact that $\tilde{H}_w\alpha^2(t+\bar{t})\leq 1-\epsilon$ because $\a \in W_\e$ according to (\ref{eqn:Wepsilon}) with $\delta < \epsilon$, means that
$\tilde{H}_x(A+BK)^{\bar{t}}\tilde{x}(t)+\tilde{H}_w\a^2(t+\bar{t})\leq 1-\delta$
implying that $((A+BK)^{\bar{t}}\tilde{x}(t),\a^2(t+\bar{t})) \in \tilde{\O}^\delta_\infty$. Hence, $t+\bar{t} \in T_a$, contradicting the assumption that $t+i \in T_b, \forall i$.
\begin{lemma}\label{lem:math}
	Consider the sequence of $a_i=\frac{\delta_i}{\beta^1_it_i+\beta^2_i}$ of (\ref{eqn:Pdiff2}) with $\delta_i>\delta>0$,  $\delta$ from (\ref{eqn:tildeOdelta}). Then $\sum_{i=1}^{\infty} a_i =\infty$.
\end{lemma}
\textbf{Proof: }
Let $\bar{\beta}^1:=\lim \sup {\beta^1_i}$ and $\bar{\beta}^2:=\lim \sup |{\beta^2_i}|$.
Then $\beta_i^1 t_i+\beta_i^2\leq \bar{\beta}^1 t_i+\bar{\beta}^2=\bar{t}(\bar{\beta}^1\frac{t_i}{\bar{t}}+\frac{\bar{\beta}^2}{\bar{t}})$. Using Lemma \ref{lem:Tb1}, $t_{i}\leq t_{i-1}+\bar{t}$, it implies $t_i\leq t_0+i \bar{t}$, or, $\frac{t_i}{\bar{t}}\leq \frac{t_0}{\bar{t}}+i$. Hence $\bar{\beta}^1\frac{t_i}{\bar{t}}+\frac{\bar{\beta}^2}{\bar{t}}\leq \bar{\beta}^1(i+\frac{\bar{\beta}^1 t_0+\bar{\beta}^2}{\bar{\beta}^1\bar{t}})$. Therefore $a_i=\frac{\delta_i}{\beta_i^1 t_i+\beta_i^2}\geq \frac{\frac{\delta}{\bar{t}\bar{\beta}^1}}{i+\frac{\bar{\beta}^1 t_0+\bar{\beta}^2}{\bar{\beta}^1\bar{t}}}:=\frac{c_1}{i+c_2}$ where $c_1:=\frac{\delta}{\bar{t}\bar{\beta}^1} >0$ and $c_2:=\frac{\bar{\beta}^1 t_0+\bar{\beta}^2}{\bar{\beta}^1\bar{t}}$  are constants.

Since $\sum_{i=1}^{\infty}\frac{c_1}{i+c_2}\ge\sum_{k=1+\bar{c}_2}^{\infty}\frac{c_1}{k}=\infty$ where $\bar{c}_2$ is upper integer bound of $c_2$. Hence,  $\sum_{i=1}^{\infty}a_i=\infty$.

\bibliographystyle{unsrtnat}
\bibliography{Reference}






\end{document}